\documentclass[aps,prd,fleqn,superscriptaddress]{revtex4}
\usepackage{graphicx,color,natbib}
\usepackage{amsmath,amssymb,amsfonts}
\newcommand{\bse}{\begin{subequations}}
\newcommand{\ese}{\end{subequations}}
\newcommand{\be}{\begin{equation}}
\newcommand{\ee}{\end{equation}}
\newcommand{\bea}{\begin{eqnarray}}
\newcommand{\eea}{\end{eqnarray}}
\newcommand{\ba}{\begin{array}}
\newcommand{\ea}{\end{array}}

\input amssym.def
\input amssym.tex

\usepackage[colorlinks=true, linkcolor=blue, bookmarks=true]{hyperref}
\begin{document}
\title{Meson Excitation Time as a Probe of Holographic Critical Point}
\author{Ali Hajilou\footnote{hajilou@mi-ras.ru}}
\affiliation{Steklov Mathematical Institute, Russian Academy of Sciences, Gubkina str. 8, 119991, Moscow, Russia}
\affiliation{School of Particles and Accelerators, Institute for Research in Fundamental Science (IPM),
P.O. Box 19395-5746, Tehran, Iran}
\affiliation{Department of Physics, Shahid Beheshti University G.C., Evin, Tehran 19839, Iran}
\begin{abstract}
 We study the time evolution of expectation value of Wilson loop as a non-local observable in  a strongly coupled field theory with a critical point at finite temperature and nonzero chemical potential, which is dual to an asymptotically AdS charged black hole via gauge/gravity duality.
 Due to inject of energy into the plasma, the temperature and chemical
potential increase to finite values and the plasma experiences an out-of-equilibrium process.
By defining meson excitation time $t_{ex}$ as a time at which the meson falls into the final excited state, we investigate the behavior of $t_{ex}$  near the critical point as the system evolves towards the critical point.
We observe that by increasing the interquark distance the dynamical critical exponent increases smoothly. Also, we obtain for slow quenches different values of the dynamical critical exponent, although for fast quenches our result for the dynamical critical exponent is in agreement with the one that is reported for studying the quasi-normal modes. 
Consequently, this indicates that in this model for fast quenches and small values of interquark distances the gauge invariant Wilson loop is a good non-local observable to probe the critical point.
 
\end{abstract}
\maketitle
\tableofcontents
\section{Introduction and results}
 Quantum Chromodynamics (QCD) that describes the strong force, is a strongly coupled gauge theory at low energy. Studying strongly coupled gauge theories has been attracted much interest. The main motivation comes from the results of Relativistic Heavy Ion Collider (RHIC) and Large Hadron Collider (LHC) that a new phase of matter, i.e. Quark Gluon Plasma (QGP) is reported \cite{CasalderreySolana:2011us,Ammon:2015wua}. At the beginning of plasma formation it is very hot, dense medium and out-of-equilibrium system. Hydrodynamic simulations show that this plasma is a strongly coupled system, so that the plasma can be realized as strongly coupled phase of QCD \cite{Shuryak:2003xe,Shuryak:2004cy}.
Thermalization process is the process of a black hole formation in the gravity theory and in the dual gauge theory side it generally means the evolution of a state from zero temperature to a thermal state  \cite{Chesler:2013lia}. An example of the thermalization process is the collapse of a thin-shell matter described by an AdS-Vaidya metric \cite{Balasubramanian:2010ce,Balasubramanian:2011ur,Bhattacharyya:2009uu}. Due to the energy injection, using an external source an out-of-equilibrium state in field theory can be produced \cite{Buchel:2014gta}. Then, one can probe the out-of-equilibrium state using local and non-local observables.
Although, lattice gauge theory describes successfully the low energy properties of QCD in particular at $\mu=0$ \cite{Rothkopf:2011db} but the lattice calculations are not successful when we include the chemical potential.
Since the plasma is strongly coupled, the perturbation theory is not applicable and we need to use a non-perturbative approach.

Anti de-Sitter/Conformal Field Theory (AdS/CFT) correspondence or more generally gauge/gravity duality is a non-perturbative approach. According to this duality, a strongly coupled gauge theory without any gravitational degrees of freedom living in a $d$-dimensional space-time is in correspondence with the classical Einstein gravity in a ($d+1$)-dimensional space-time \cite{Maldacena:1997re,Witten:1998qj,Gubser:1998bc}. In fact, parameters and different processes in the gauge theory side is translated into the corresponding equivalent in the gravity side. Utilizing this duality, many different problems in the strongly coupled regime of QCD are investigated \cite{Gursoy:2007cb,Gursoy:2007er,Dudal:2018ztm}. For instance,
in order to calculate the static potential energy between a quark and an anti-quark in the strongly coupled plasma we need to calculate expectation value of a Wilson loop. In fact, Wilson loop is non-local and gauge invariant observable that for the first time in \cite{Maldacena:1998im} is used to calculate the potential energy between a pair and its generalization is extensively discussed in
\cite{Brandhuber:1998bs,Rey:1998bq,Sonnenschein:2000qm,Finazzo:2013aoa,Ali-Akbari:2015ooa,Dudal:2017max,Cai:2012xh,Bohra:2019ebj,Asadi:2021nbd}.
 Gauge/gravity duality proposes that the holographic dual of the rectangular Wilson loop is given by a classical open string suspended from two points on the boundary of the gravity and hanging down in the gravity bulk with appropriate boundary conditions. In addition, this duality is used to study various aspects of strongly coupled systems such as jet quenching parameter \cite{Liu:2006ug,Cai:2012eh}, thermalization process \cite{Galante:2012pv,Ali-Akbari:2012gku,Ali-Akbari:2012fzl,Dey:2015poa}
 and the features of plasma in presence of magnetic field 
\cite{Ali-Akbari:2015bha,Bohra:2020qom,Ali-Akbari:2013txa,Arefeva:2020vae,Zhou:2020ssi,Dudal:2021jav,Arefeva:2021mag,Fang:2021ucy,Dudal:2016joz,He:2020fdi}.
For more details, see \cite{CasalderreySolana:2011us} and references therein.
 
Understanding the phase structure of QCD is received a lot of interest \cite{Qin:2010nq}.
The endpoint of the line of first-order phase transition is described as a critical point.
 In particular, studying the physics of  observables near the critical point in QCD phase diagram is challenging question. 
It is very important to note that the investigation of this question is difficult theoretically because the theory is strongly coupled near the critical point\cite{DeWolfe:2010he}. To do so, the gauge/gravity duality prepared a new approach in such a way that to investigate a strongly coupled theory near the critical point one can consider the  black hole solution which is holographically dual to a strongly coupled field theory with a critical point \cite{Ebrahim:2017gvk,Cai:2022omk}.
 
  In this paper, we consider  Einstein-Maxwell-dilaton (EM-dilaton) background and its Vaidya-like solution dual to a field theory with a critical point \cite{Zhang:2015dia,Ebrahim:2017gvk}. In fact, in the gauge theory side  we consider a probe stable meson in the QGP at zero temperature and chemical potential. Due to the injection of energy the temperature and the chemical potential are increased from zero to finite values of $T$ and $\mu$, respectively. A very interesting observation in  \cite{Hajilou:2018dcb} is the time that a meson bound state needs to fall into the final excited state is called \textit{excitation time}, $t_{ex}$.
 Now the question we would like to answer is whether the meson excitation time, $t_{ex}$, can probe the critical point as the system evolves towards the critical point? Furthermore, what would be the associated dynamical critical exponent? Holographic critical point and dynamical critical exponent have been extensively discussed in \cite{DeWolfe:2010he,DeWolfe:2011ts}. The holographic dual of the above system in the gravity side is described by the dynamics of the classical open string with appropriate initial and boundary conditions in the Einstein-Maxwell-dilaton-Vaidya (EM-dilaton-Vaidya) background.
 
 Our main findings can be summarized as follows:

\begin{itemize}
\item
We observed that by moving towards the critical point the excitation time, $t_{ex}$, gets the finite value thought its slope, $\frac{dt_{ex}}{d\frac{\mu}{T}}$, diverges at the critical point and its behavior can be described by $((\frac{\mu}{T})_{\star}-\frac{\mu}{T})^{-\theta}$ where $(\frac{\mu}{T})_{\star}$ indicates the valus of $\frac{\mu}{T}$ at the critical point and $\theta$ is defined as the dynamical critical exponent.

\item
Our observation showed the dynamical critical exponent $\theta$ depends on the interquark distance $l$ (or dimensionless quantity $lT$) and the speed of energy injection $k$ that is whether the quench is fast or slow. 

 \item
We observed that increasing both values of $k$ and $l$  the value of $\theta$ increases.
Also, it is seen that the effect of changing $k$ on $\theta$ is much more than the effect of  $l$ on $\theta$ and we emphasize that this feature is our observation. 
It seems that, except for the non-locality of our probe some features of our results can be addressed to the non-equilibrium conditions that is set up in the system. 

\item
A very interesting observation is that for fast quenches ($k\ll 1$) and small values of interquark distances $l$, the dynamical critical exponent is $\frac{1}{2}$ which is in good agreement with the result that is obtained from the investigation of the behavior of scalar quasi-normal modes near the critical point in \cite{Finazzo:2016psx}. Therefore, in this model the  gauge invariant Wilson loop is good non-local observable to probe the critical point.

\item
 Another point is that the dependence of the dynamical critical exponent $\theta$ to the transition time $k$ is also reported in \cite{Ebrahim:2017gvk}. 
 Although, there is an important difference here since our physical observable, Wilson loop, is a non-local operator versus of the local operator  that is used in \cite{Ebrahim:2017gvk}.
Also, because of non-locality of Wilson loop it is anticipated that the value of $\theta$  depends to the $l$. This is because the dependence of meson excitation time, $t_{ex}$, to $l$ is confirmed in \cite{Hajilou:2018dcb} and therefore $\theta$ depends on $l$.

\end{itemize}

 The remainder of this paper is organized as follows. In section \ref{two} we briefly review on the EM-dilaton-Vaidya background  and then calculate the time evolution of expectation value of Wilson loop in this background.
In Section \ref{three} we explain our numerical results. 
 Finally, in Section \ref{outloook} we discuss our results and compare with previous researches.
 This work is complemented by Appendix \ref{c1}, where we review on the EM-dilaton black hole and the EM-dilaton-Vaidya  backgrounds. In Appendices \ref{a1} and \ref{b1} we obtain  appropriate boundary and initial conditions, respectively. In Appendix \ref{numeric} we introduce our numerical procedure to obtain the dynamical critical exponent. In Appendix \ref{zero17} we considered meson potential at zero temperature and in Appendix \ref{stringplot22} we depicted 3-dimensional plot of open string that describes the time evolution of classical open string in EM-dilaton-Vaidya background.

\section{ Probing the critical point by a non-local observable} \label{two}
Gauge/gravity duality establishes a good framework to study important features of QGP \cite{CasalderreySolana:2011us,Ammon:2015wua}.
Studying the phase structure of gauge theories is a long-standing project and one of the curiosities in this field is how the system behaves near the critical point.  In other words, utilizing local or non-local observables the critical point can be probed and one can also calculate  the associated dynamical critical exponent.
To do so, in this section we review a charged black hole solution which its holographic dual is a field theory with a critical point  and  also a non-local observable, i.e. Wilson loop that is used to probe the critical point in the field theory.

\subsection{ Background}
We want to study the evolution of expectation value of Wilson loop in the time-dependent and strongly coupled field theory with a critical point. Its holographic dual is to study the dynamics of an open string in the EM-dilaton-Vaidya background (for more details see Appendix \ref{c1}). The  metric in Eddington-Finkelstein coordinates is given by \cite{Zhang:2015dia}:
\be\label{metric120}\begin{split}%
 ds^2=&-N(z) F(\bar{v},z) d\bar{v}^2 -\frac{2}{z^2}\sqrt{\frac{N(z)}{1+b^2z^2}} d\bar{v}dz+ \frac{1+b^2z^2}{z^2}g(z)d\vec{x}^2,\\
F(\bar{v},z)=&\frac{1+b^2z^2}{z^2}\Gamma^{2\gamma}-M(\bar{v})\frac{z^2}{1+b^2z^2}\Gamma^{1-\gamma},
\end{split}\ee%
 where,
\be \begin{split} \label{metric130}
N(z)=\Gamma^{-\gamma},~g(z)=\Gamma^{\gamma},~\Gamma(z)=\frac{1}{1+b^2z^2},~\gamma=\frac{\alpha^2}{2+\alpha^2}~,
\end{split} \ee
$\alpha$ is the coupling constant between the dilaton and the gauge field and $b$ is a constant.
The $\bar{v}$-coordinate reduces to the time coordinate of the gauge theory at the boundary, i.e. $t=\bar{v}|_{z=0}$. $M(\bar{v})$ and $q(\bar{v})$ are arbitrary functions that are used to show how the mass and  charge of the  black hole increase from zero to a finite value. Different functional forms of $M(\bar{v})$ and $q(\bar{v})$ are discussed in the literature, for example in \cite{Amiri-Sharifi:2016uso}. Here, our choice is
\bea %
 {\zeta}(\bar{v})=  {\zeta} \left\{%
\begin{array}{ll}
0 & {\bar{v}}<0, \\
k^{-1}\left[{\bar{v}}-\frac{k}{2\pi}\sin(\frac{2\pi {\bar{v}}}{k})\right] & 0 \leqslant {\bar{v}} \leqslant k, \\
1 & {\bar{v}}>k ~,\\
\end{array}%
\right.
\eea
where ${ \zeta} \in(M , q)$. The time interval that the black hole needs to reach its final value of the mass and charge, i.e. $M$ and $q$, respectively is the transition time $k$. In the gauge theory side $k$ is the time that strongly coupled system needs to reach its final values of temperature and chemical potential. The {\it{fast (slow) quench}} is attributed to the $k\ll 1$ ($k\gg 1$), that is small (large) transition time, respectively. 
The relation between the final values of mass and charge of the black hole, i.e. $M$ and $q$  and the constant $b$ are:
\be\label{parameter} 
q=\sqrt{\frac{6M}{2+{\alpha}^2}}b ~.
\ee
It is very important to note that for studying the phase structure of the gauge theory we consider $\alpha=2$ in this background.
 After fixing the value of $\alpha=2$ the equation (\ref{parameter}) indicates that this background is parameterized by three parameters $q$, $M$, $b$, while two of them are independent.
The chemical potential of the strongly coupled field theory, that comes from the gauge field in the bulk is given by \cite{Zhang:2015dia}:
\be \label{mu10}
\mu=\frac{b\sqrt{3M}}{(\frac{1}{{z_h}^2}+b^2)\sqrt{2(2+{\alpha}^2)}}~,
\ee
Where the $z_h$ is the horizon radius and is the smallest root of the equation $f(z_h)=0$ (see Appendix \ref{c1}). 

\subsection{Thermodynamics}
 
\subsubsection{Temperature and entropy}

The temperature of the field theory that is correspondence with the Hawking temperature of the black hole is \cite{Zhang:2015dia,Ebrahim:2017gvk}:
\be \label{mu11}
T=\frac{b{\Gamma(z_h)^{\frac{3\gamma}{2}-1}}}{4\pi\sqrt{1-\Gamma(z_h)}}[2(3\gamma-1)-3(2\gamma-2)\Gamma(z_h)]~.
\ee
After simplification of the temperature formula in terms of parameters of the theory we obtained:
\be\label{tem19} 
T=\frac{1}{z_h}~\frac{3+\sqrt{1+4~ q^2 z_h^6}}{2 \pi \sqrt{2}~ \sqrt{1+\sqrt{1+4~ q^2 z_h^6}}}~.
\ee
Also, the entropy (density) can be written as:
\be\label{entropy15} 
s=\frac{1}{z_h^3}~\frac{\sqrt{1+\sqrt{1+4~ q^2 z_h^6}}}{4~\sqrt{2} }~.
\ee
It is important to note that by choosing the case $\alpha =2$, this background possesses a critical point at $(\frac{\mu}{T})_{\star}$=1.11072. One can find more details and its phase diagram in \cite{Ebrahim:2017gvk}. To do so, utilizing equations (\ref{mu10}) and (\ref{mu11}) we have:
\be\label{parameter10} 
bz_h=\frac{1 \pm \sqrt{1-\frac{8{\mu}^2}{\pi^2 T^2}}}{\frac{2 \mu}{\pi T}}~,
\ee
where 
\be\label{parameter11} 
z_h=\sqrt{\frac{b^2+\sqrt{b^4+4m}}{2m}}~.
\ee
 \begin{figure}[ht]
\begin{center}
\includegraphics[width=66 mm]{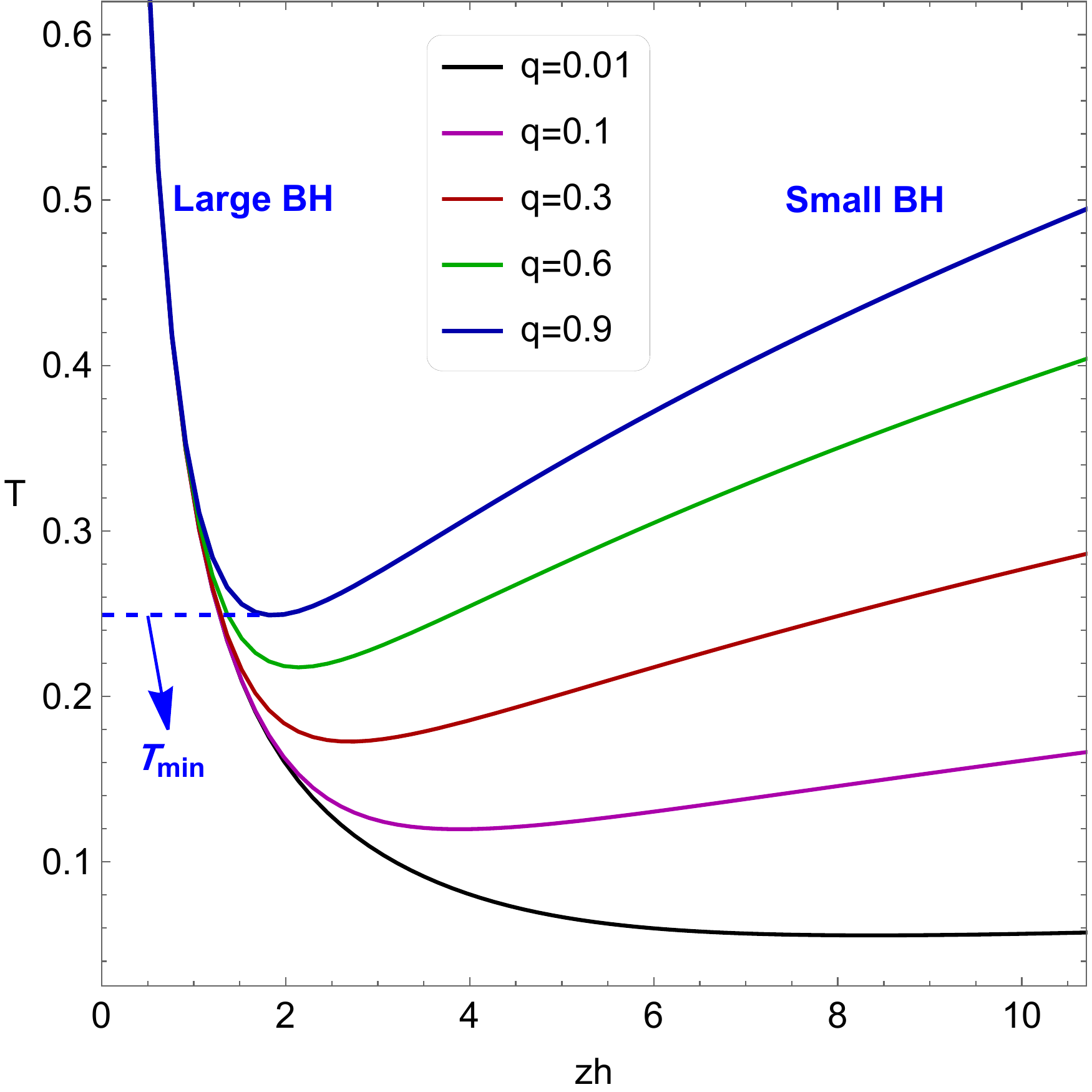}
\hspace{13 mm}
\includegraphics[width=66 mm]{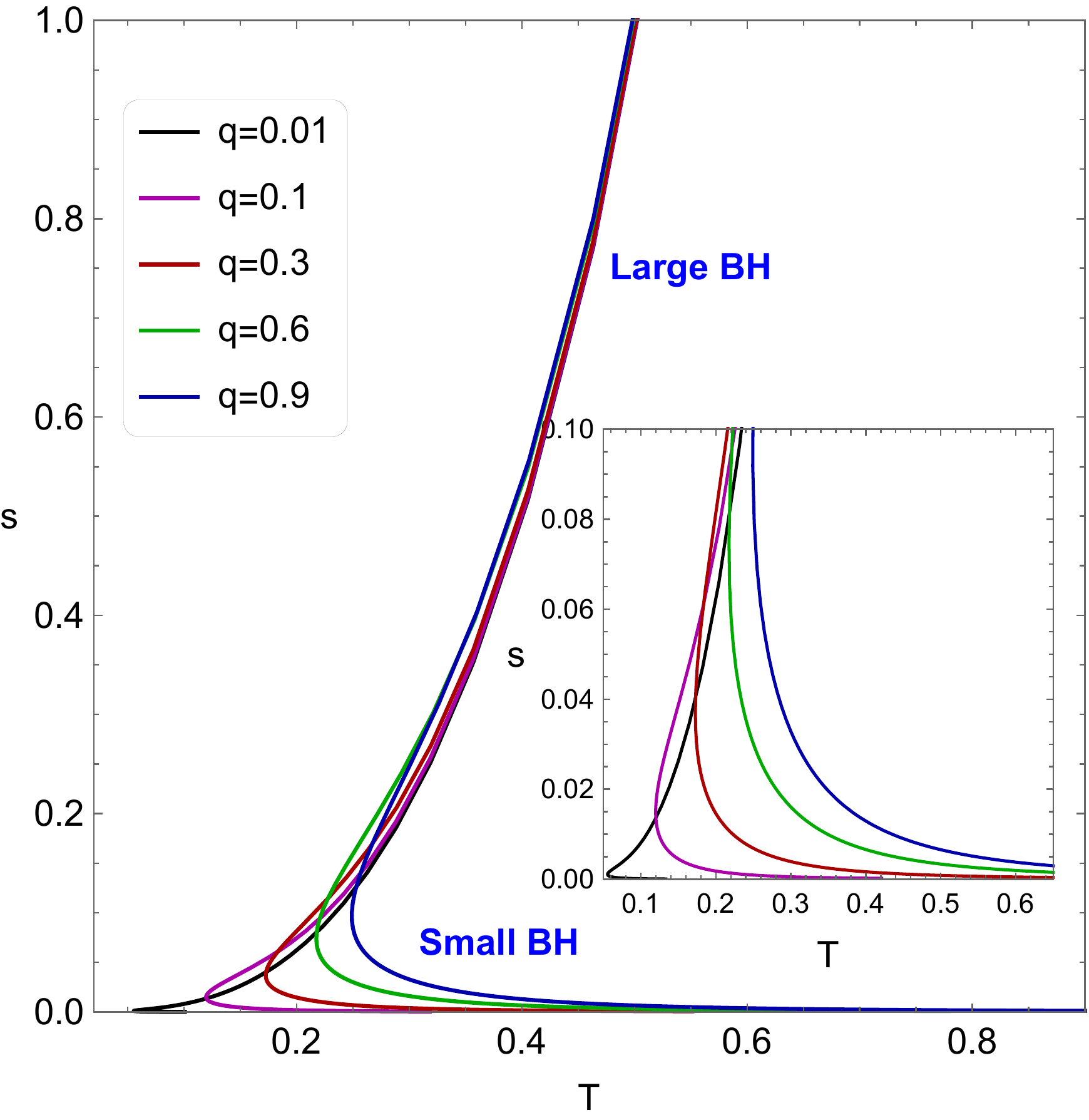}
\caption{ Left: Temperature of the black hole $T$ as a function of the horizon $z_h$ for different values of charges. 
Right: Entropy of the black hole $s$ as a function of the temperature $T$ for different values of charges. For each panel, their values of charges have been fixed as $q=0.01$ (Black),  $q=0.1$ (Magenta), $q=0.3$ (Red), $q=0.6$ (Green), $q=0.9$ (Blue).
\label{free2}}
\end{center}
\end{figure}
It is obvious that for each value of $\frac{\mu}{T}$, there are two distinct values of $bz_h$ which describes stable and unstable branches of solutions.  In other words, the lower sign in equation (\ref{parameter10}) corresponds to thermodynamically stable solution and the upper sign, unstable. Consequently, this indicates that there exist a phase transition in field theory and where the two branch of solutions merge, there exists a \textit{critical point} \cite{Ebrahim:2017gvk}.
In fact, in this theory the black hole solutions possess the "critical point" at which the thermodynamical stability of a black hole solution switches.
 Note also that one can check the  thermodynamically stable or unstable solutions utilizing the Jacobian, $\mathcal{J}=\frac{\partial (s,\rho)}{\partial (T,\mu)}$
where $s$ is the entropy and $\rho$ is the charge density
\be \label{srho} \begin{split}%
 s &\propto \frac{T^3(1+b^2 z_h^2)^2}{(2+b^2 z_h^2)^3}  ~,\\
 \rho &\propto  \frac{\mu}{T}(2+b^2 z_h^2) \sqrt{1+b^2 z_h^2}~.
\end{split} \ee%
In fact, if the Jacobian is positive (negative), there exists a thermodynamically stable (unstable) solution for the physical system \cite{Finazzo:2016psx}. In this research  the thermodynamically stable solution is used.

 Studying the thermodynamics of this background, i.e. EM-dilaton black hole can be helpful to investigate the black hole features and phase transition.
In figure \ref{free2}, the variation of Hawking temperature $T$ with respect to the horizon radius $z_h$ is depicted for different values of black hole charges, i.e. $q=0.01$ (Black),  $q=0.1$ (Magenta), $q=0.3$ (Red), $q=0.6$ (Green), $q=0.9$ (Blue). We observe that there exists a minimum temperature $T_{min}$ in such a way that below
which there is no black hole solution. However, for $T>T_{min}$, there are two black hole solutions, large black hole and small one that are specified in the figure \ref{free2}. The large black hole that its temperature increases when $z_h$ decreases, is stable, whereas the small black hole that its temperature increases when $z_h$ increases, is unstable one.  This plot also shows that as the charge of the black hole $q$ increases, the minimum temperature increases. 
Note also that, the stable-unstable solutions can be investigated also by studying the entropy of the black hole. In the right panel of the figure \ref{free2}, we depicted the behavior of entropy $s$ as a function of the temperature $T$ for the same values of black hole charges. As shown in the right panel, there is a minimum of the temperature such that for the temperature larger than minimum, we have two different black hole solutions. At the one hand, the solution that its entropy increases when temperature increases is stable (large black hole) and on the other hand,  the solution that its entropy increases when temperature decreases is unstable one (small black hole). We have clarified these two branches of solutions in the right panel of the figure \ref{free2}. It is important to note that the stable-unstable nature of the large-small black hole phases can be seen easily from the free energy behavior that is investigated in the figure \ref{free3}. 
  
 
 \subsubsection{Free energy and phase transition}
 
\begin{figure}[ht]
\begin{center}
\includegraphics[width=68 mm]{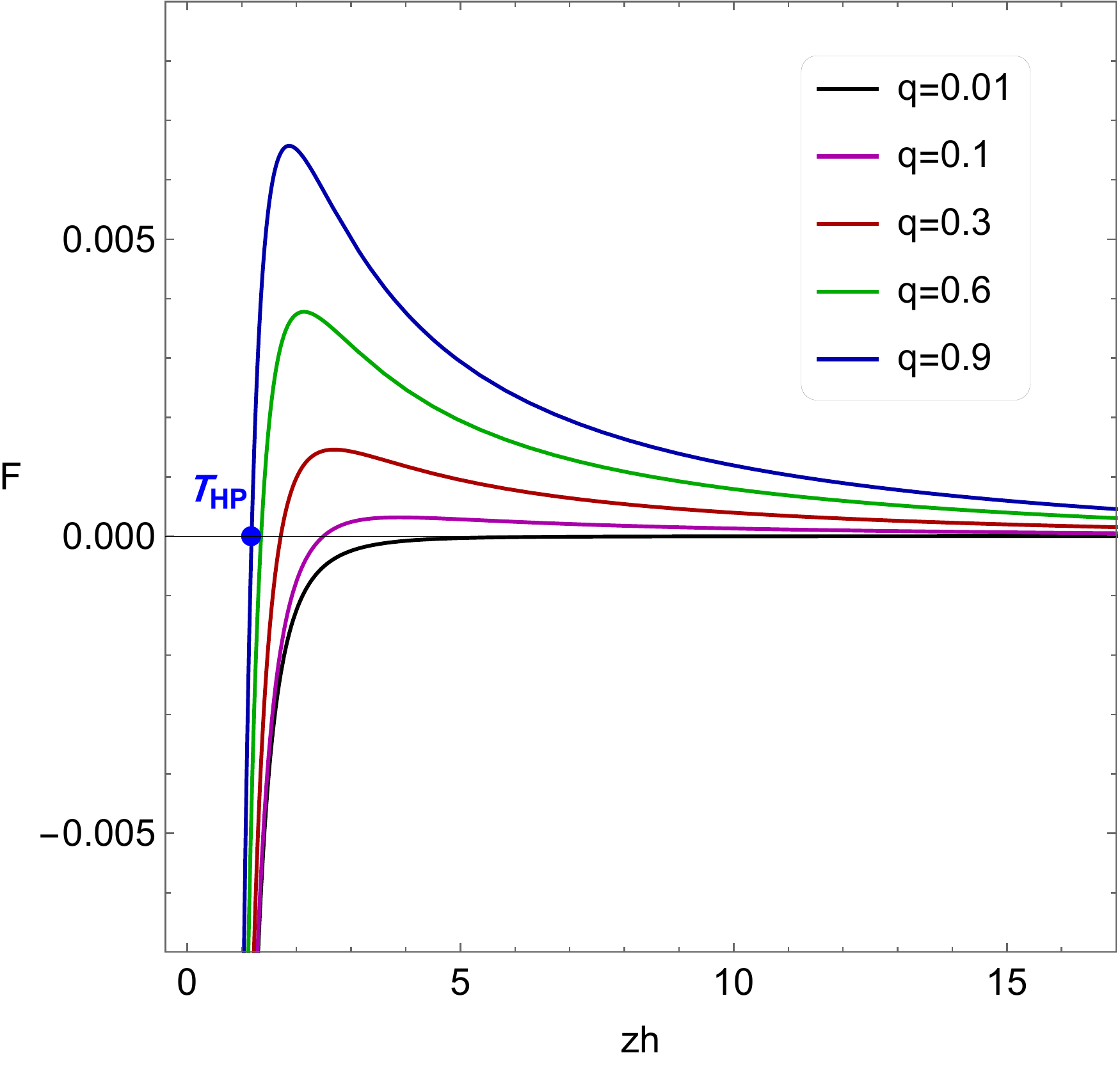}
\hspace{13 mm}
\includegraphics[width=68 mm]{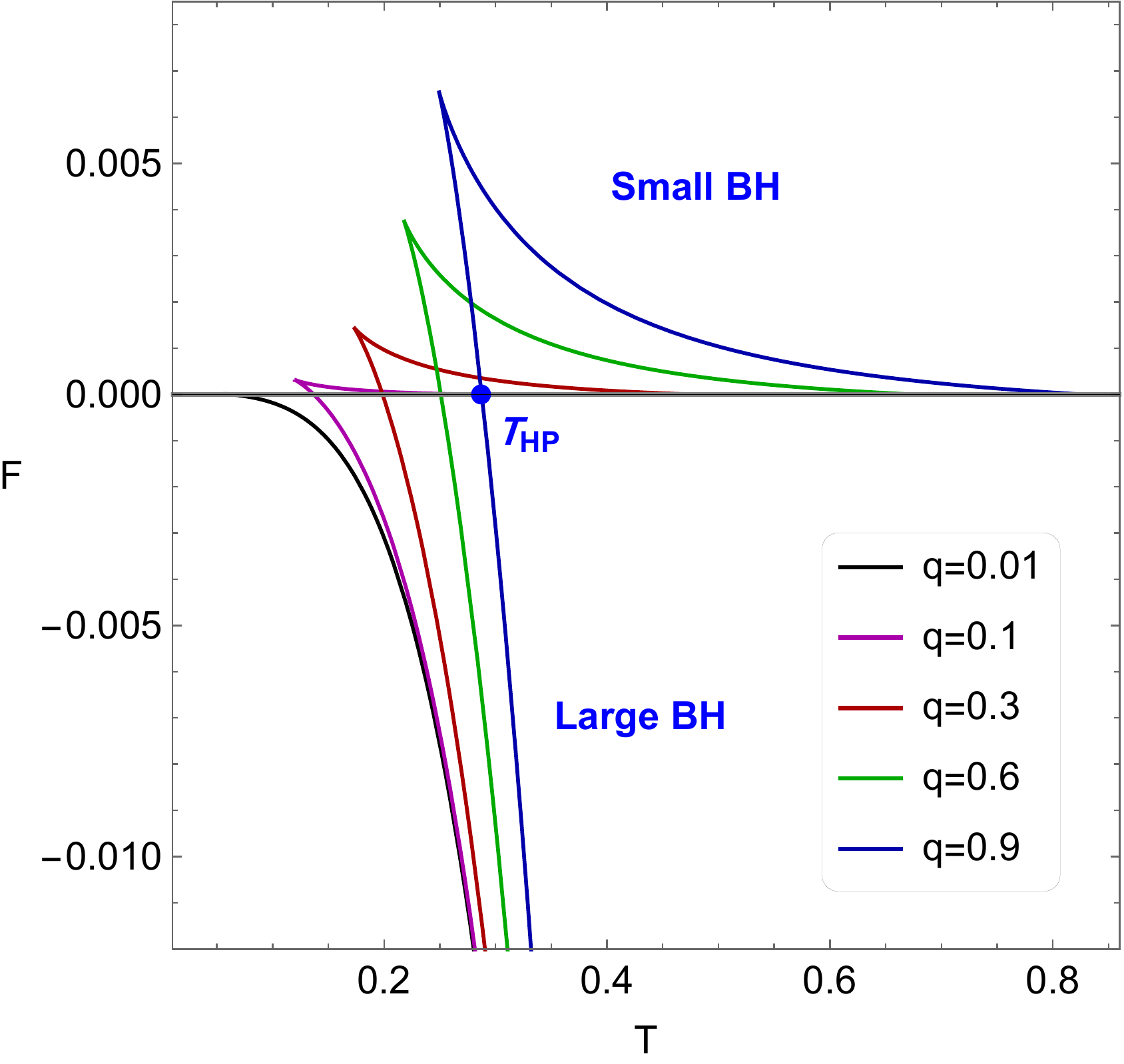}
\caption{  Left: Free energy of the black hole $F$ as a function of the horizon $z_h$ for different values of charges. We have normalized the free energy such that it vanishes when $z_h \rightarrow \infty $.
Right: Free energy of the black hole $F$ as a function of the temperature $T$ for different values of charges. For each panel, values of charges have been fixed as $q=0.01$ (Black),  $q=0.1$ (Magenta), $q=0.3$ (Red), $q=0.6$ (Green), $q=0.9$ (Blue). 
\label{free3}}
\end{center}
\end{figure}

Free energy is very powerful physical quantity to investigate the stability and unstability of the solutions. According to the first law of thermodynamics the free energy (density) is defined \cite{DeWolfe:2010he,Arefeva:2020vae,He:2013qq}:
\be\label{free100}
F=\epsilon - sT- \mu \rho ~,
\ee 
where $\epsilon$ is energy density and $\rho$ is number density. The differential of free energy at fixed volume is obtained as $dF = - sdT- \rho d\mu $. For the fixed values of chemical potential $\mu$, the free energy can be obtained by the following integral \cite{Gursoy:2008za,Arefeva:2018hyo,Gursoy:2018ydr}:
\be\label{free102}
dF = - \int sdT ~.
\ee 
We expect that at $z_h \rightarrow \infty $ the free energy of the black hole back ground, coincides with the free energy of the zero temperature back ground (thermal gas) that can be choose to be zero. Therefore, we have normalised the free energy of the black hole with respect to the thermal gas case by demanding $F(z_h \rightarrow \infty)=0 $ \cite{Bohra:2019ebj,He:2013qq} . Then, one can obtain the following relation for the free energy:
\be\label{free102}
F = \int_{z_h}^{\infty} s~\frac{dT}{dz_h} dz_h ~.
\ee 
The behavior of free enregy $F$ in terms of horizon radius $z_h$ is shown in the left panel of figure \ref{free3} for different values of black hole charges, $q=0.01$ (Black),  $q=0.1$ (Magenta), $q=0.3$ (Red), $q=0.6$ (Green), $q=0.9$ (Blue). As shown in figure \ref{free3} we observe that the sign of free energy changes. In other words, at small $z_h$  the free energy gets large negative value and then gets positive maximum value and finally decreases to zero at $z_h \rightarrow \infty $. The free energy intersecting the horizontal axis implies that there exists a phase transition from the black hole to the thermal gas. It is more transparent to investigate the behavior of free energy $F$ in terms of temperature $T$ as depicted in the right panel of figure \ref{free3}.  We observe that the free energy of the small black hole phase is always larger than the large black hole and thermal gas phases. This indicates the unstable nature of the small black hole phase. It is important to note that, upon varying the Hawking temperature, a phase transition from the large black hole phase to thermal AdS phase takes place at the Hawking-Page transition temperature $T_{HP}$. This is the famous black hole-thermal AdS (Hawking-Page) phase transition \cite{Hawking:1982dh} that is in correspondence with confinement-deconfinement phase transition and is usually known as a first order phase transition in the field theory \cite{Arefeva:2018hyo,He:2022amv,Li:2017tdz}.

\begin{figure}[ht]
\begin{center}
\includegraphics[width=82 mm]{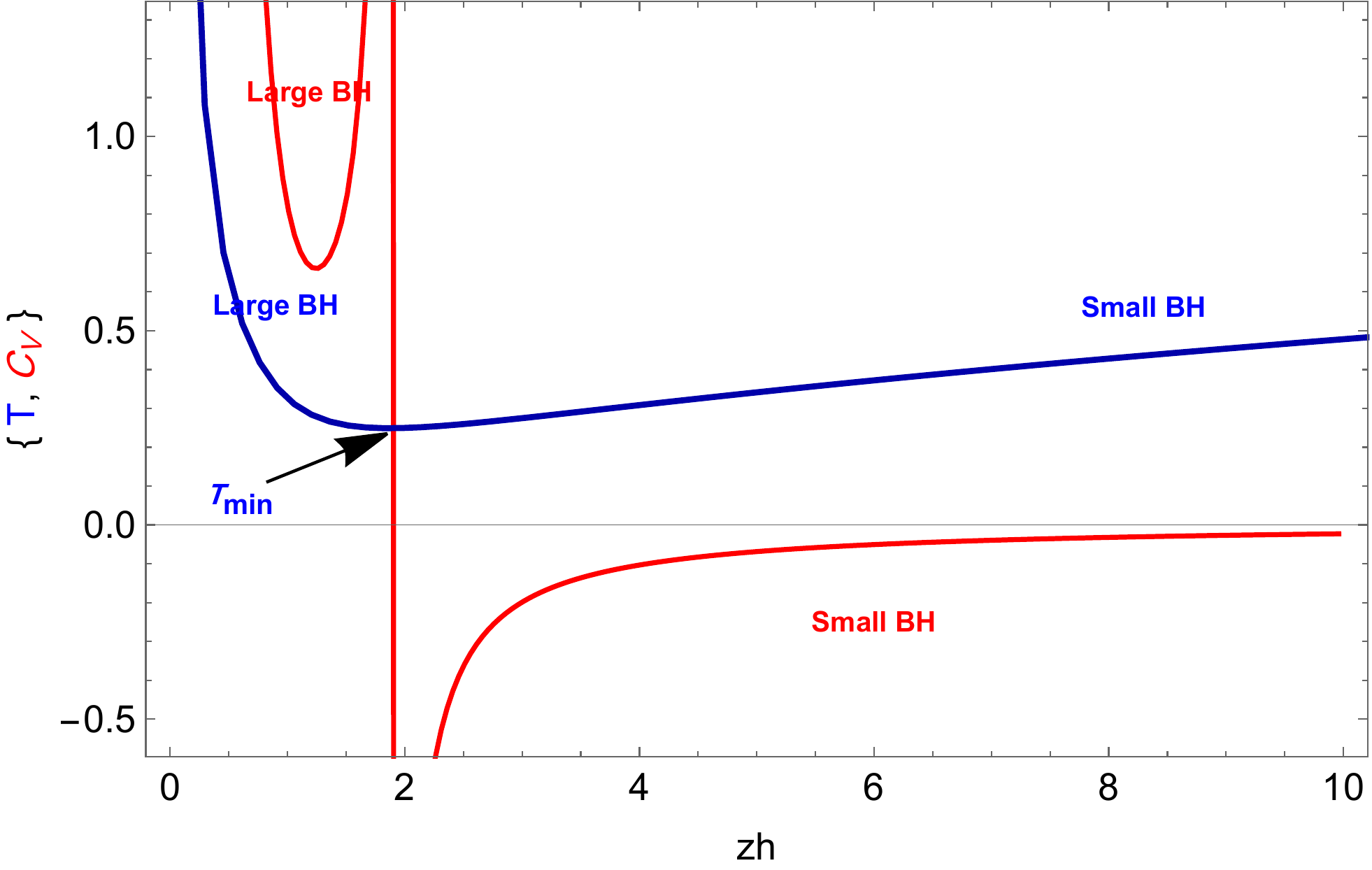}
\hspace{10 mm}
\includegraphics[width=80 mm]{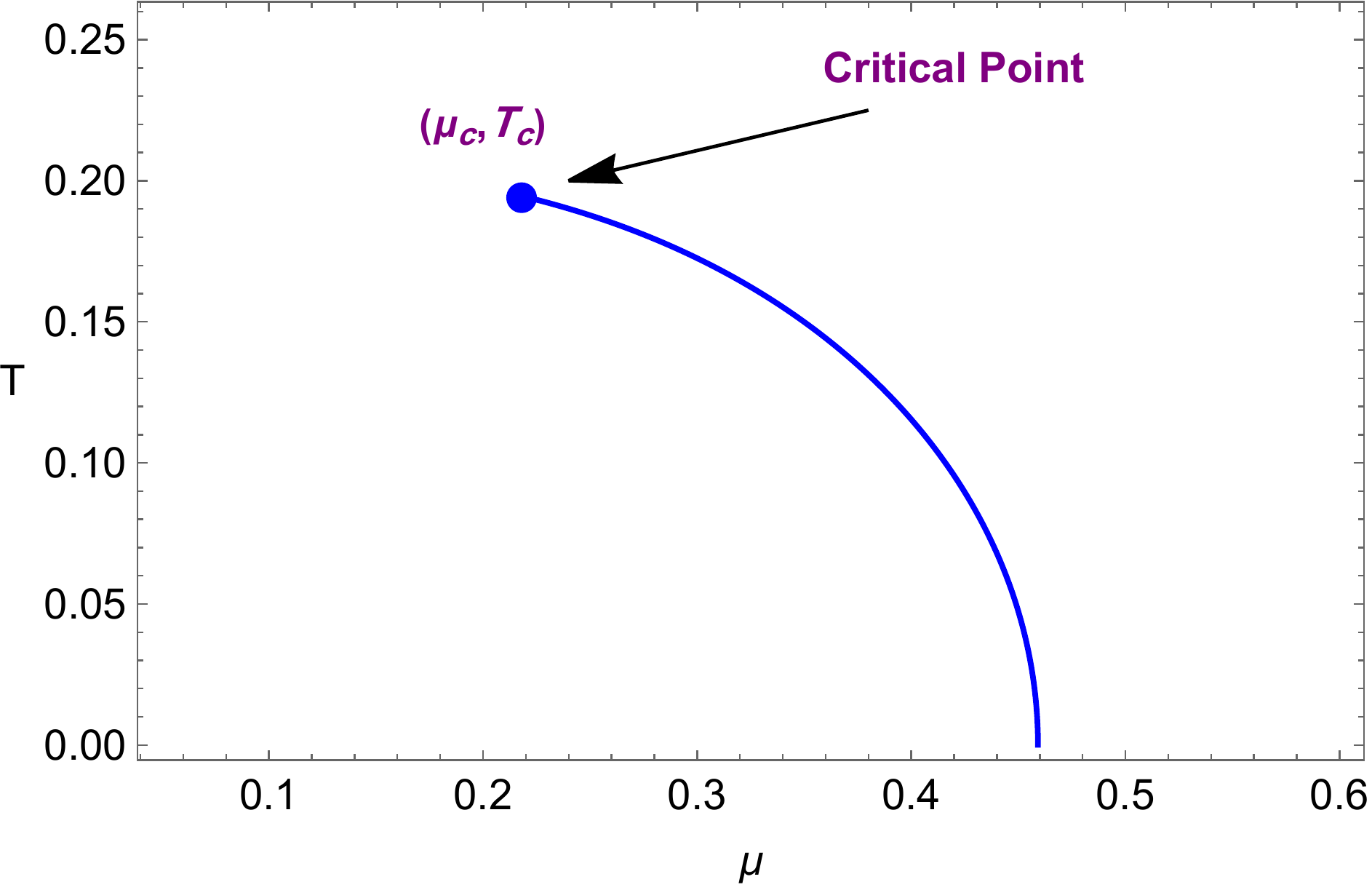}
\caption{  Left: Temperature $T$ (blue line) and heat capacity $C_V$ (red line) of the black hole as a function of the horizon radius $z_h$. For both plots, we have fixed the value of charge as $q=0.9$.
Right: Temperature of the black hole $T$ as a function of the chemical potential $\mu$. The critical point $(\mu_c , T_c)$ is shown in this panel such that we have $\frac{\mu_c}{T_c} \sim 1.1107$ . 
\label{free4}}
\end{center}
\end{figure}

 The critical point is a point that the Jacobian $\mathcal{J}$ vanishes. The Jacobian $\mathcal{J}$ is proprtional to the heat capacity $C_V$ of the system, i.e.  $\mathcal{J} \sim C_V$ \cite{DeWolfe:2010he}. Note also that, for heat capacity at constant volume we have \cite{DeWolfe:2010he}:
 \be\label{free108}
 C_V=T\left(\frac{\partial s}{\partial T}\right)_V ~.
\ee
Therefore, we depicted in the left panel of figure \ref{free4} the heat capacity of the system $C_V$  (red line) in terms of the horizon radius $z_h$ to investigate where the Jacobian vanishes (changes its sign) \cite{He:2013qq,Sajadi:2023zke} and the temperature $T$ (blue line) as a function of horizon radius $z_h$. A very interesting  observation in the left panel of figure \ref{free4} is that the point where the Jacobian changes its sign (changing from stable solution to the unstable one), is in coincidence with the $T_{min}$ (where we have changes from stable branch of solution to the unstable one). It is important to note that, although in this model the free energy just could  show the confinement-deconfinement phase transition and was not capable to illustrates critical point $T_{crit}$ (since we have no Swallow-tail diagram), but, using the left panel of figure \ref{free4} we could describe the coincidence of the point where the stable and unstable solutions transforms to each other. In fact, this observation tells us that Jacobian is not good order parameter to see Critical End Point (CEP) in this model.
Note also that, in the right panel of figure \ref{free4} we plotted the temperature $T$ as a function of chemical potential $\mu$ via the formula:
 \be\label{free109}
 \mu=\frac{\sqrt{1+\pi T zh (-\pi T zh +\sqrt{-1+\pi^2 T^2 zh^2 })} }{\sqrt{2}~ zh } ~.
\ee
This panel shows the first order phase transition line that the end of this line is the CEP with $(\mu_c , T_c)$ where we have $\frac{\mu_c}{T_c} \sim 1.1107$. 

\subsection{ Wilson loop} 

Wilson loop is a gauge invariant and non-local observable that is very useful to find the potential energy between quark and anti-quark pair, living in the plasma. 
The holographic dual of the rectangular Wilson loop is given by a two-dimensional world-sheet swept by a classical open  string suspended from two points (corresponding to a quark and an anti-quark), hanging down in the bulk  with appropriate boundary conditions.
 We used the time-like rectangular Wilson loop $\cal{C}$, where one side of the loop is  the spatial distance between quark and anti-quark pair, $l$, and the other side is the temporal direction, $\cal{T}$. If we suppose that ${\cal{T}}\gg l$, that is the world-sheet is translationally invariant along the time direction, the expectation value of the Wilson loop is \cite{CasalderreySolana:2011us}:
\be\label{wilson3}
\langle W({\cal{C}}) \rangle =e^{-i(2 m + V(l)){\cal{T}}},
\ee
where $m$ is the rest mass of quark (antiquark) that is equal to $\frac{\sqrt{\lambda}}{2\pi}\int_{\epsilon}^{z_h}\,\frac{dz}{z^2}$, where $\epsilon$ is IR regulator in the gravity theory that according to the UV/IR connection corresponds to the UV cut-off in the gauge theory side \cite{CasalderreySolana:2011us}. Also, $V(l)$ represents static potential energy between the pair. According to AdS/CFT dictionary, the expectation value of the Wilson loop, in the saddle point approximation, is dual to the on-shell action of the two-dimensional world-sheet of an open string whose dynamics is given by Nambu-Goto action. Therefore,
\be\label{wilson4}
\langle W({\cal{C}})\rangle =e^{i S({\cal{C}})},
\ee\label{action}
where $S({\cal{C}})$ is the Nambu-Goto action: 
\be\label{action2}
S=\frac{-1}{2 \pi \alpha'} \int d\tau d\sigma \sqrt{- \det(g_{ab})}\,,
\ee
$\tau$ and $\sigma$  parametrize the two dimensional world-sheet of the string and ${\alpha}^\prime={l_s}^2$ and ${l}_s$ is the fundamental length scale of string. The induced metric on the string world-sheet is $g_{ab}=G_{\mu\nu}\frac{\partial X^\mu}{\partial \xi^a} \frac{\partial X^\nu}{\partial \xi^b}$ so that, $G_{\mu\nu}$ is the bulk metric. Here, $X^\mu$ ($\xi^a=\tau,\ \sigma$) are the bulk (world-sheet) coordinates. Using equations (\ref{wilson3}), (\ref{wilson4}) and (\ref{action2}) the static potential energy between the pair can be found.

In order to calculate the Nambu-Goto action in the EM-dilaton black hole background (\ref{metric1}) (see Appendix \ref{c1}), we work in static gauge and choose $\tau= t$, $\sigma=x_3\equiv x$ to parametrize the two-dimensional string world-sheet. Therefore, except $z$ and $x$, the other bulk coordinates are chosen to be constant and the function $z=z(x)$ describes the shape of the classical string. Therefore, the action (\ref{action2}) on background (\ref{metric1}) reduces to
\be\label{staticaction}
S=-\frac{{\cal{T}}}{2\pi\alpha'}\int_{-\frac{l}{2}}^{\frac{l}{2}}\, dx\frac{\sqrt{N(z)}}{z^2}
\sqrt{\frac{{z^\prime}^2}{1+b^2 z^2}+z^2{f}(z)g(z)(1+b^2 z^2)}~,
\ee
where $z'\!=\!dz/dx$. Since the Lagrangian does not depend explicitly on $x$, the associated Hamiltonian is a constant of motion and will be used later to obtain initial conditions (see Appendix \ref{b1}). After some simple algebra, one gets
\be\label{staticz}
z'(x)=\pm\sqrt{\frac{N(z)}{N(z_\ast)}}\frac{g(z)f(z)~z_\ast}{\sqrt{g(z_\ast)f(z_\ast)}}\sqrt{\frac{(1+b^2 z^2)^3}{(1+b^2 z_\ast^2)}}~
\sqrt{1- \frac{z^2~f(z_\ast)g(z_\ast)N(z_\ast)(1+b^2 z_\ast^2)}{z_\ast^2 ~f(z)g(z)N(z)(1+b^2 z^2)}}~,
\ee
where $z=z_\ast$ is where $z'(x)=0$. 

For studying the evolution of Wilson loop in the time-dependent plasma, one should inject energy into the plasma. Since the system is time dependent, it is not translationally invariant along the time direction. Therefore, the condition ${\cal{T}}\gg l$ will not work as before and therefore the expectation value of the Wilson loop, (\ref{wilson3}) in time-dependent case can be written as \cite{Ali-Akbari:2015ooa,Hajilou:2018dcb,Hajilou:2017sxf}
\be\label{timewilson}
\langle W({\cal{C}})\rangle =e^{-i \int dt\ {\cal{W}}(t)}\,.
\ee%
Based on the gauge/gravity duality, ${\cal{W}}(t)$ is the on-shell action of string where the integration over time coordinate has not been done. In order to regularize $ {\cal{W}}(t)$, we subtracted the infinite mass of the quark and anti-quark pair. Thus, we have
\be\begin{split}\label{regul}%
{\cal{W}}_R(t) = {\cal{W}}(t) - 2 m \equiv \int d\sigma \left(\sqrt{- \det (g_{ab})}\right)_{\rm{on-shell}} - 2 m~,
\end{split}\ee%
where ${\cal{W}}_R(t) $ is the regularized form of ${\cal{W}}(t)$ and describes the time dependence of expectation value of Wilson loop. In order to calculate ${\cal{W}}_R(t) $ in the gauge theory side, we need to calculate the on-shell action of the string in the EM-dilaton-Vaidya background.

 To do so, similar to \cite{Ishii:2014paa,Hajilou:2017sxf,Hajilou:2018dcb}, to parametrize the two-dimensional world-sheet of the string, we chose the null coordinates $(u,v)$ on the world-sheet. Therefore, all the background coordinates on the world-sheet depend on $u$ and $v$ and all coordinates will be zero except the following ansatz:
\be\label{ansatz7}%
\bar{v}=V(u,v)~ , ~ z=Z(u,v) ~,~ x_3=X(u,v)\,.
\ee%
Substituting this ansatz into the Nambu-Goto action (\ref{action2}), the equations of motion can be obtained. After some algebra we found
\be \label{eom}\begin{split}%
X_{,uv}& = \left(Z_{,u} X_{,v} + Z_{,v} X_{,u}\right)\frac{3+2b^2Z^2}{3Z(1+b^2Z^2)} ~, \\
V_{,uv}&=
\left(\frac{2b^2Z^3 F }{3}(1+b^2Z^2)^{\frac{-1}{6}}+\frac{Z^2 F_{,Z}}{2}(1+b^2Z^2)^{\frac{5}{6}}\right) V_{,u} V_{,v}
\\ 
&+\left(\frac{1}{Z}(1+b^2Z^2)^{\frac{1}{2}}-\frac{b^2Z}{3}(1+b^2Z^2)^{\frac{-1}{2}}\right) X_{,u} X_{,v} ~, 
\\ 
%
Z_{,uv}&=\left(-\frac{2b^2F^2Z^5}{3}(1+b^2Z^2)^{\frac{2}{3}} -\frac{FZ^4F_{,Z}}{2}(1+b^2Z^2)^{\frac{5}{3}}-\frac{Z^2F_{,V}}{2}(1+b^2Z^2)^{\frac{5}{6}} \right) V_{,u} V_{,v}
\\ 
&+ \left(-\frac{2b^2Z^3F}{3}(1+b^2Z^2)^{\frac{-1}{6}}-\frac{Z^2F_{,Z}}{2}(1+b^2Z^2)^{\frac{5}{6}}\right) \left(Z_{,u} V_{,v}+Z_{,v} V_{,u}\right)
\\ 
&+ \left( -Z F(1+b^2Z^2)^{\frac{4}{3}} +\frac{b^2Z^3F}{3}(1+b^2Z^2)^{\frac{1}{3}}\right) X_{,u} X_{,v}+
\left( \frac{2}{Z}+\frac{b^2Z}{3}(1+b^2Z^2)^{-1}\right) Z_{,u} Z_{,v} ~.
%
\end{split}\ee%
Since $u$ and $v$ are null coordinates, two constraint equations corresponding to $g_{uu}=0$ and $g_{vv}=0$ should be imposed. Therefore, we have
\be\label{cons}\begin{split}%
C_1 &= Z^2F(V,Z) V_{,u}^2 + 2(1+b^2Z^2)^{\frac{-5}{6}} V_{,u}  Z_{,u} - (1+b^2Z^2)^{\frac{-1}{3}}X_{,u}^2  = 0~,
 \\
C_2& = Z^2 F(V,Z) V_{,v}^2 + 2(1+b^2Z^2)^{\frac{-5}{6}} V_{,v}  Z_{,v} - (1+b^2Z^2)^{\frac{-1}{3}} X_{,v}^2  = 0 \,.
\end{split}\ee%
In order to solve the equations of motion (\ref{eom}) and constraint equations (\ref{cons}), we need to impose suitable boundary and initial conditions. In Appendices \ref{a1} and \ref{b1}, the appropriate boundary and initial conditions are obtained, respectively.

\section{Numerical results} \label{three}
Having set up the formalism developed in the previous section, now we are ready to
 discuss the numerical results in this section . Consider a probe stable meson in the QGP at zero temperature and chemical potential. Then, after the injection of energy into the plasma, the temperature and the chemical potential raised to final values, i.e. $T$ and $\mu$, respectively. Before the injection of energy the meson is in its ground state and the injection of energy puts the meson into a final excited state with specific frequency and amplitude of oscillation \cite{Ali-Akbari:2015ooa,Hajilou:2018dcb,Ishii:2014paa,Ageev:2016gtl}. The time that the meson needs to fall into the final excited state is called \textit{excitation time}, $t_{ex}$ \cite{Hajilou:2018dcb}. In other words, the excitation time, $t_{ex}$, is the time that the expectation value of the Wilson loop starts oscillating around the static potential energy by which we mean the potential energy of the bound state in the plasma with final values of the temperature and chemical potential $T$ and $\mu$, respectively. 
An important point that we would like to emphasize is that since we work in the probe limit, therefore the energy of the meson does not dissipate in the plasma and consequently the oscillation of meson remains unchanged.

\begin{figure}[ht]
\begin{center}
\includegraphics[width=80 mm]{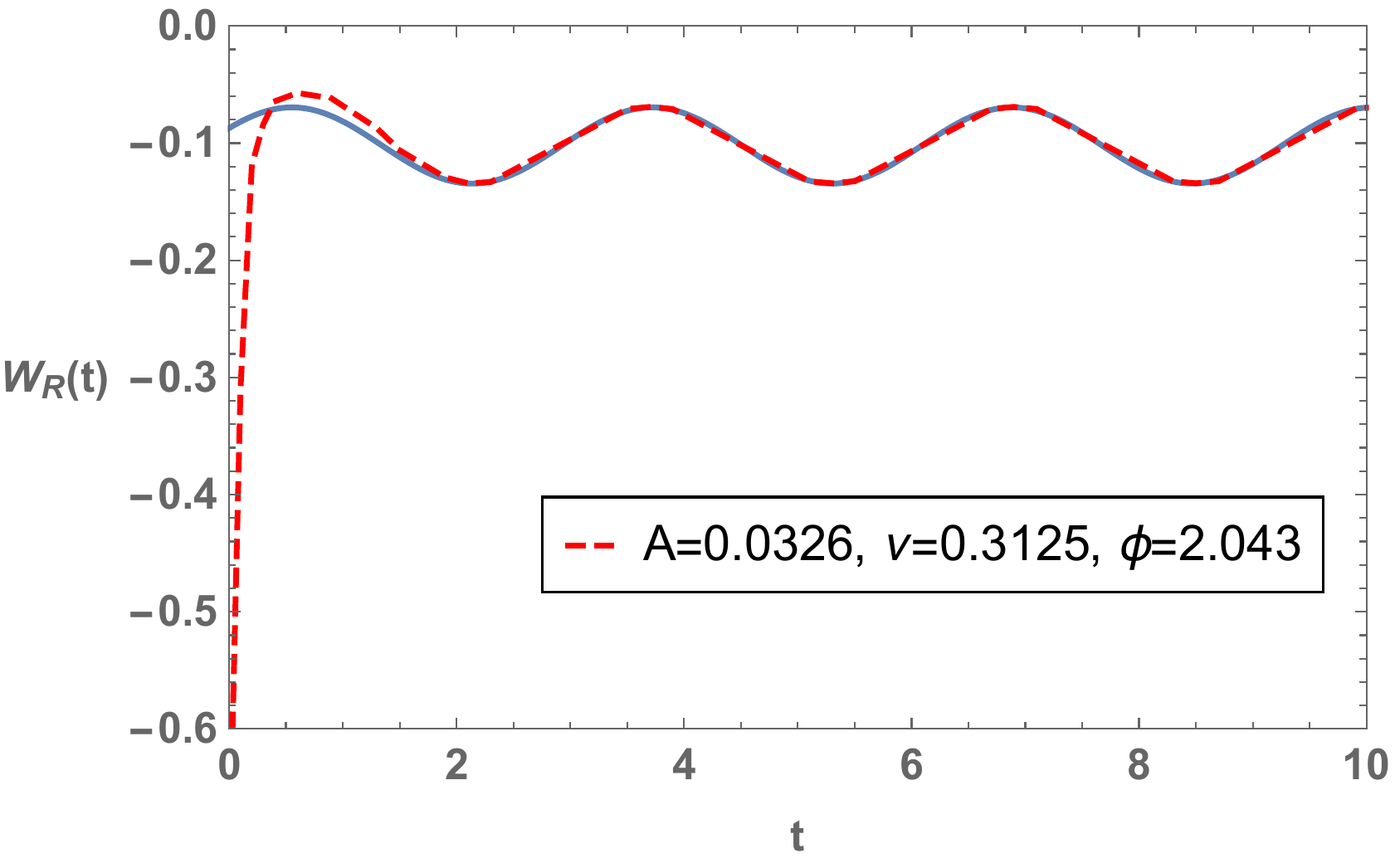}
\caption{Evolution of the expectation value of Wilson loop ${\cal{W}}_R(t)$ as a function of boundary time $t$. We fixed the interquark distance $l=1$, final value of the chemical potential $\mu=0.0220$, final value of the temperature $T=0.2200$ and the transition time $k=0.3$. The dashed red curve shows the evolution of Wilson loop and the blue sine curve is the fitted function, i.e. ${\cal{W}}_f(t) = A\cos (2 \pi \nu t +{\phi})$ where, $A$ is the amplitude of the oscillation, $\nu$ is the oscillation frequency and $\phi$ is a phase value. Reprinted from \cite{Hajilou:2018dcb}. 
\label{ex11}}
\end{center}
\end{figure}

Note that the response of the system to the time-dependent change in the temperature and the
chemical potential is described by the time evolution of the expectation value of Wilson loop ${\cal{W}}_R(t)$. Therefore, for better clarification of the excitation time we depicted the ${\cal{W}}_R(t)$ as a function of boundary time $t$ in figure \ref{ex11}. In this figure we fixed the interquark distance $l=1$, final value of the chemical potential $\mu=0.0220$, final value of the temperature $T=0.2200$ and the transition time $k=0.3$. The value of the static potential is $V(l)=-0.1017$. As shown in figure \ref{ex11}  the expectation value of the time-dependent Wilson loop oscillates around the static potential energy.
 In fact, when the energy injection is started, the temperature and the chemical potential of the system increase and  the quark and anti-quark pair, or equivalently quark–anti-quark bound state, is excited. After finishing the energy injection, the pair falls into a final excited state and starts to oscillate with particular frequency and amplitude of oscillation.
The dashed red curve is the time evolution of the expectation value of Wilson loop that is fitted with the blue curve
\begin{equation}%
\label{fit2}
{\cal{W}}_f(t) = A\cos (2 \pi \nu t +{\phi}) \ ,
\end{equation}%
where $A$, $\nu$ and $\phi$ can be fixed from ${\cal{W}}_R(t)$ at asymptotic times. Consequently, the excitation time, $t_{ex}$, is a time that the expectation value of Wilson loop oscillates around its static potential energy with a specific value of frequency $\nu$, amplitude of oscillation $A$ and a phase value  
$\phi$. 
To cover this concept we define  a time-dependent function
\begin{equation}\label{fit11} 
{\epsilon}(t) = \mid\frac{{\cal{W}}_R(t) - {\cal{W}}_f(t)}{{\cal{W}}_R(t)}\mid \ .
\end{equation}%
Therefore, the excitation time, $t_{ex}$, is defined when the condition $\epsilon(t_{ex})<5\times 10^{-6}$ satisfies and $\epsilon(t)$ in (\ref{fit11}) remains below afterwards. For more details see \cite{Hajilou:2018dcb}.

The main important question that we are interested in is to investigate whether the meson excitation time understands about the phase structure of the gauge theory. Note that we want to know if the meson excitation time can probe the critical point. To put in another way, we study the behavior of  meson excitation time, $t_{ex}$, near the critical point when the system moves towards the critical point. 

\begin{figure}[ht]
\begin{center}
\includegraphics[width=77 mm]{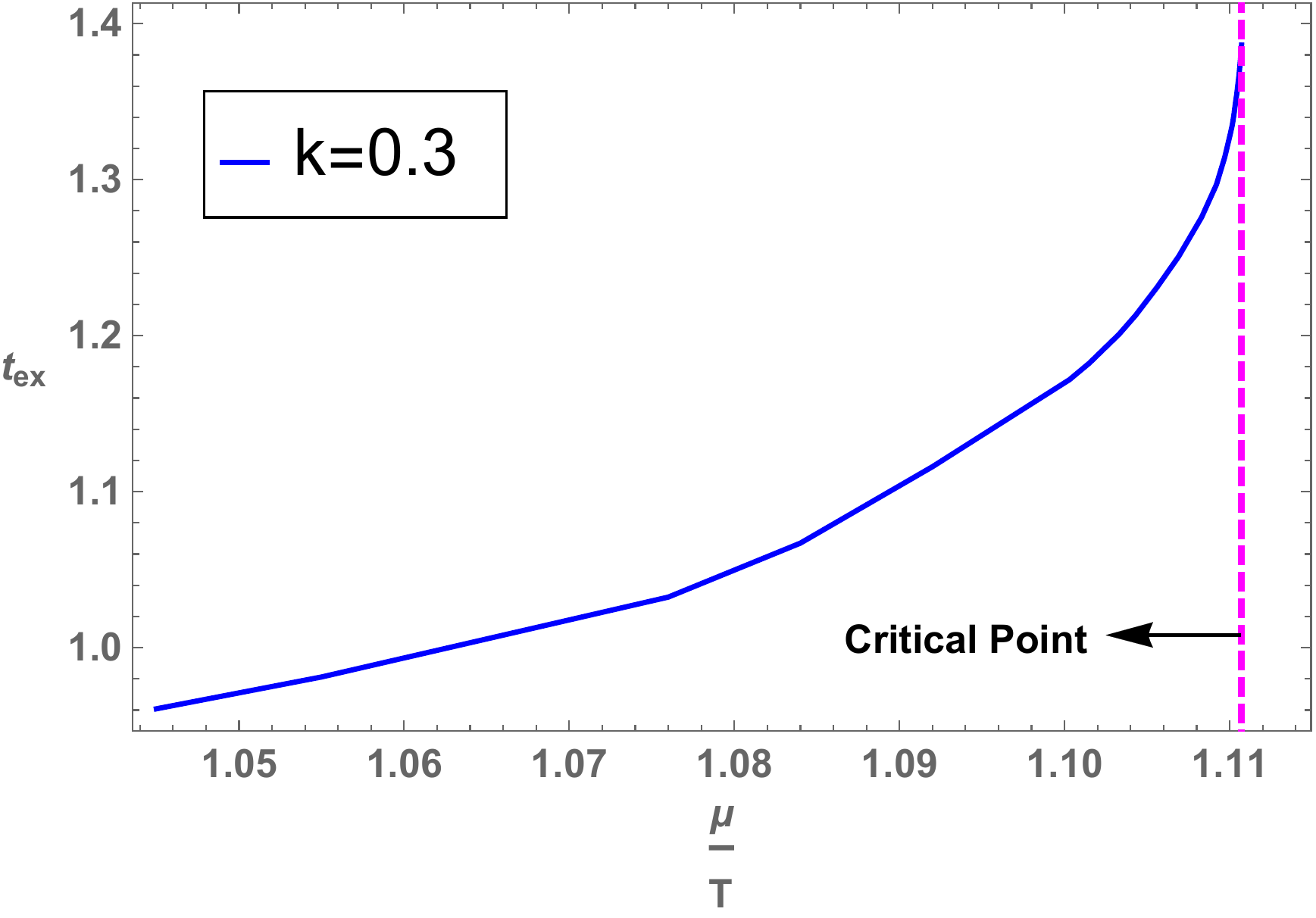}
\hspace{10 mm}
\includegraphics[width=77 mm]{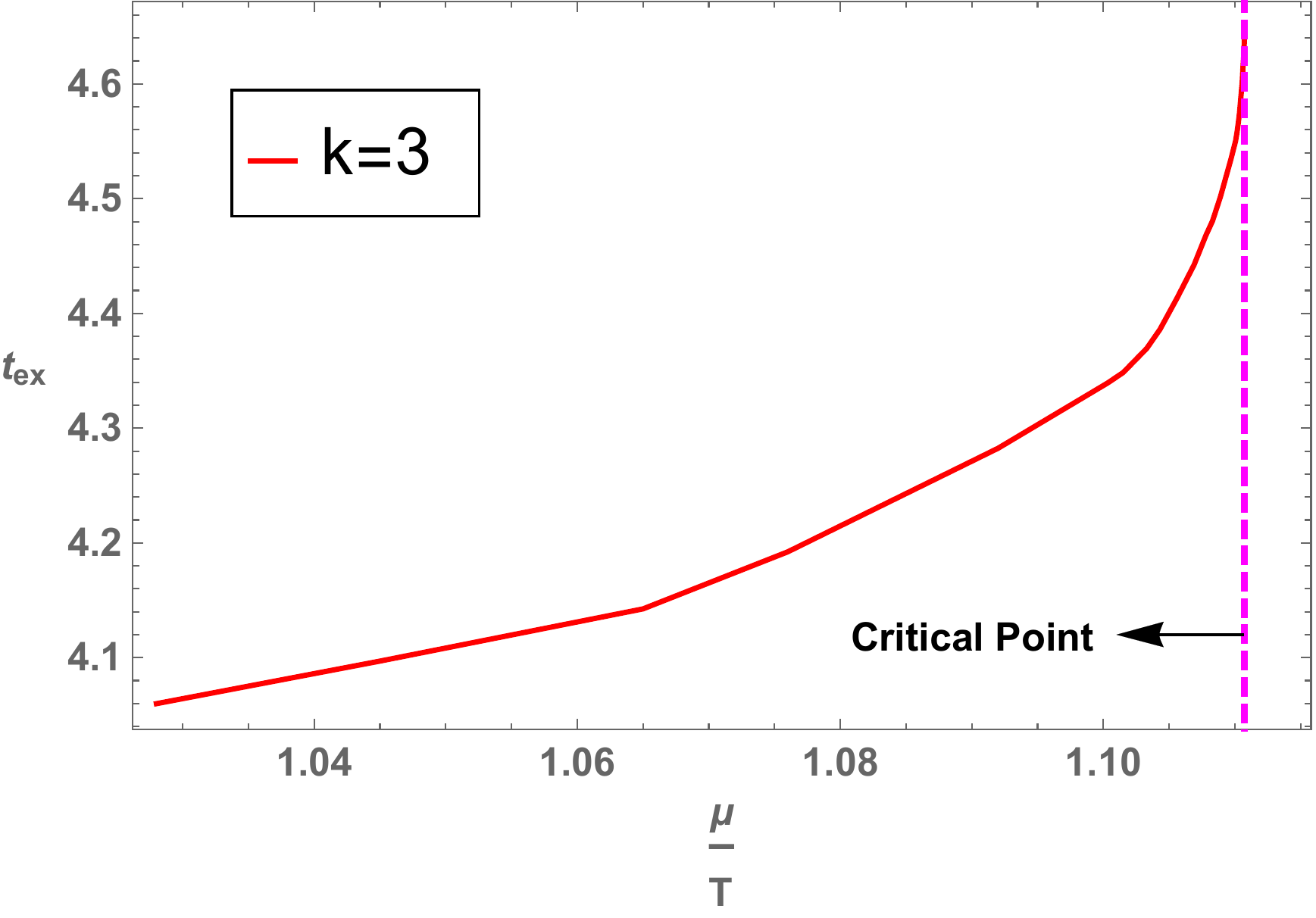}
\caption{ Excitation time $t_{ex}$ as a function of $\frac{\mu}{T}$. 
Left: We fixed the values of interquark distance $l=0.27$, the transition time $k=0.3$  and the temperature $T=0.37$ ($lT=0.10$).
Right: We fixed the values of interquark distance $l=0.27$, the transition time $k=3$  and the temperature $T=0.37$ ($lT=0.10$).
 In both left and right panels the magenta dashed line corresponds to the critical point which is at $(\frac{\mu}{T})_{\star}$=1.11072.
\label{f17}}
\end{center}
\end{figure}

In figure \ref{f17} we plotted the excitation time, $t_{ex}$, as a function of $\frac{\mu}{T}$ for fixed values of $lT=0.10$ , the temperature $T=0.37$ and  interquark distance $l=0.27$. We fixed the transition time $k=0.3$ and $k=3$ for left and right panel, respectively. In fact, 
in order to introduce the critical point and its general features for different quenches,
 we investigated the behavior of  $t_{ex}$ as a function of $\frac{\mu}{T}$ for  different quenches  $k=0.3$ ($k=3$) that corresponds to fast (slow) quench for the left (right) panel, respectively.  In both left and right panels the magenta dashed line corresponds to the critical point which is at $(\frac{\mu}{T})_{\star}$=1.11072.
 As shown in figure \ref{f17}, we observed that at the critical point, $(\frac{\mu}{T})_{\star}$,  the excitation time, $t_{ex}$, gets the \textsf{finite value} though its slope, $\frac{dt_{ex}}{d\frac{\mu}{T}}$, approaches \textsf{infinity} at this point. 
In order to check let's define the slope 
\begin{equation}\label{cc}
\frac{dt_{ex}}{d\frac{\mu}{T}} (i)=\frac{t_{ex}(i+1)-t_{ex}(i)}{\frac{\mu}{T}(i+1)-\frac{\mu}{T}(i)}~,
\end{equation}
where $i$ represents the ith point of the corresponding data points. 
It is important to note that, in our numerical results by varying two independent input parameters, we plotted the $t_{ex}$ and its slope with respect to the $\frac{\mu}{T}$. In practice, for $\frac{\mu}{T}$ we utilized
\begin{equation}
\frac{\mu}{T}=\frac{\pi \,q\,z_h \sqrt{\frac{z_h^4 +\sqrt{z_h^8+4\,q^2\, z_h^2}}{2}}}{q^2+2\,zh^2 \left(\frac{z_h^4 +\sqrt{z_h^8+4\,q^2\,z_h^2}}{2} \right)}~,
\end{equation}
where, $q$ and $z_h$ are two independent input parameters. 
A very interesting observation is that the slope of data points near the critical point can be fitted with the function
\begin{equation}
\frac{dt_{ex}}{d\frac{\mu}{T}}=(\frac{\pi}{2\sqrt{2}}-\frac{\mu}{T})^{-\theta}~,
\end{equation}
where $\theta$ is a positive number and defined as a dynamical critical exponent \cite{Ebrahim:2017gvk}. 

\begin{figure}[ht]
\begin{center}
\includegraphics[width=77 mm]{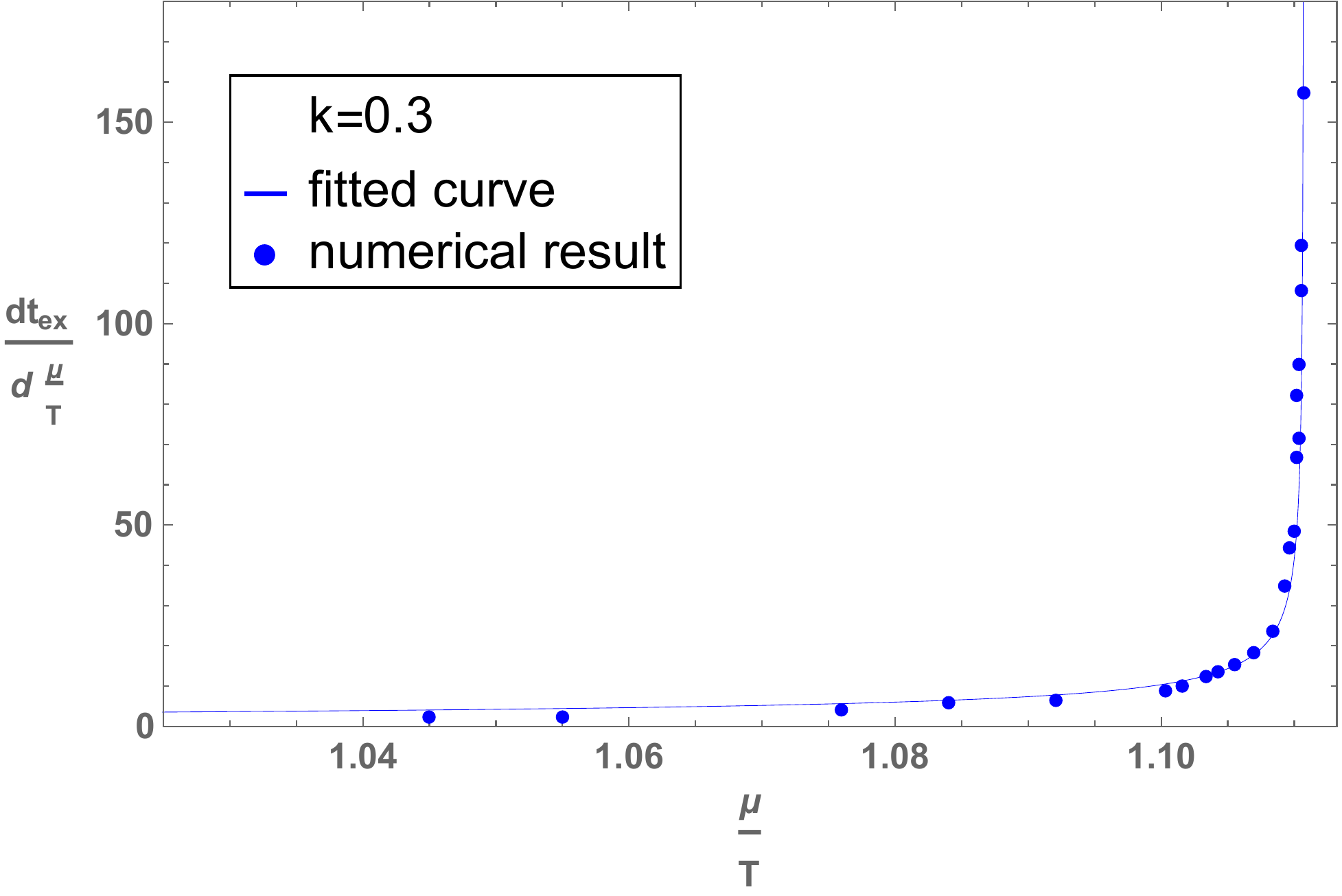}
\hspace{10 mm}
\includegraphics[width=77 mm]{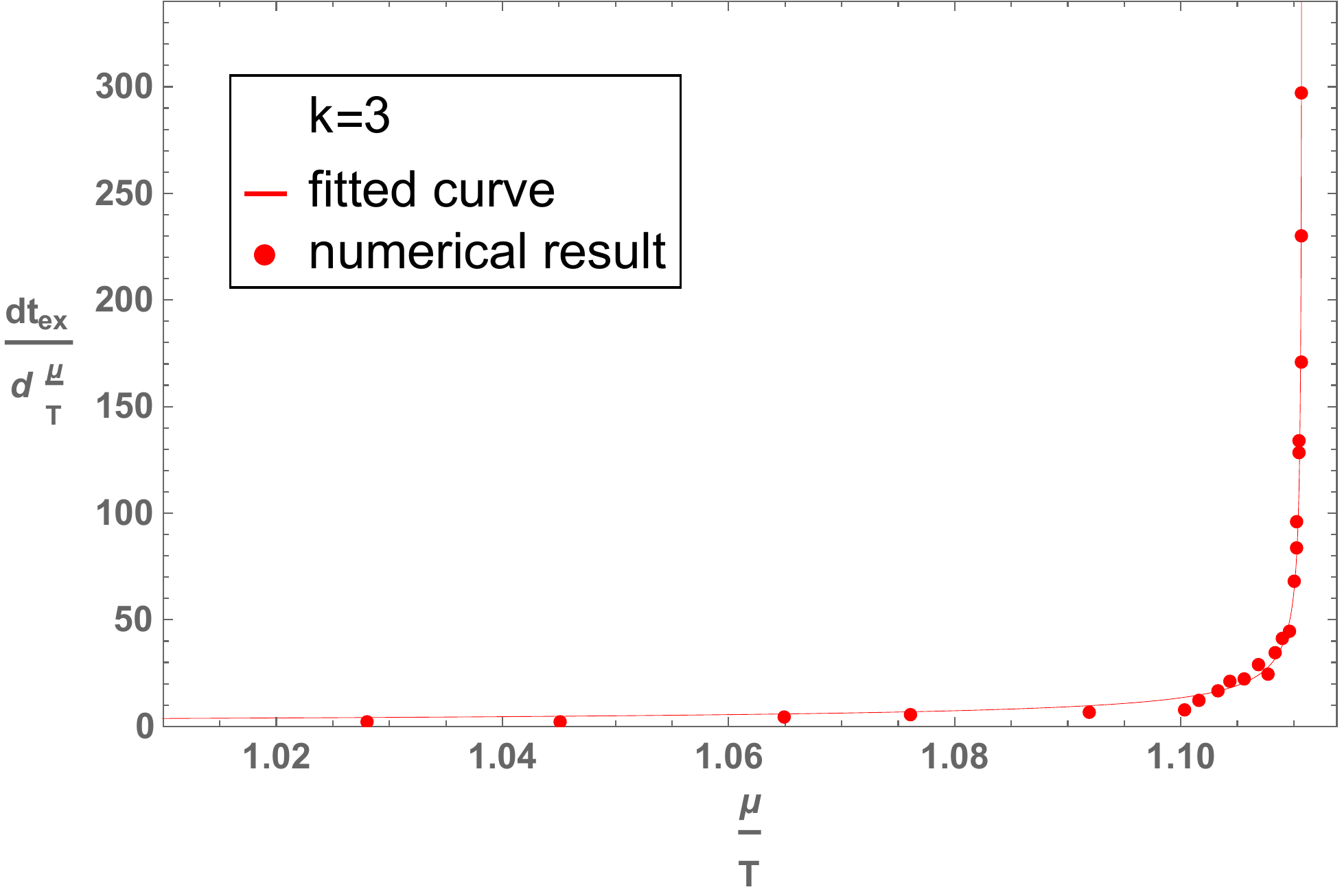}
\caption{ $\frac{dt_{ex}}{d \frac{\mu}{T}}$ as a function of $\frac{\mu}{T}$.
Left: We fixed the values of interquark distance $l=0.27$, the transition time $k=0.3$  and the temperature $T=0.37$ ($lT=0.10$). The blue curve is the function, $(\frac{\pi}{2\sqrt{2}}-\frac{\mu}{T})^{-\theta}$, fitted with the data with $\theta=0.515667$.
Right: We fixed the values of interquark distance $l=0.27$, the transition time $k=3$  and the temperature $T=0.37$ ($lT=0.10$). The red curve is the same function fitted with the data with $\theta=0.572473$. 
\label{f1}}
\end{center}
\end{figure}

As you can see in figure \ref{f1} we depicted $\frac{dt_{ex}}{d \frac{\mu}{T}}$ as a function of $\frac{\mu}{T}$ that is the slope (\ref{cc}) near the critical point. In the left and right panel, we fixed the values of interquark distance $l=0.27$ and the temperature $T=0.37$ ($lT=0.10$). But, in the left panel  we fixed the transition time $k=0.3$ that corresponds to fast quench and in the right panel we fixed the transition time $k=3$ that corresponds to slow quench.
 We observed that for the fast quench the value of dynamical critical exponent is $\theta=0.515667$ and for slow quench $\theta=0.572473$. A very interesting observation is that by increasing the value of transition time from $k=0.3$ to $k=3$, there is a smooth deviation in dynamical critical exponent that is  the dynamical critical exponent is sensitive to the value of the transition time $k$, although smoothly. Another point is that our result is in good agreement with the the result that is obtained from the investigation of the behavior of scalar quasi-normal modes near the critical point in \cite{Finazzo:2016psx}.
 Consequently, the gauge invariant and non-local observable, i.e. Wilson loop is  a good observable to probe the critical point of the theory when the system evolves towards the critical point. 
 Recently, different observables is investigated to find dynamical critical exponent \cite{Ebrahim:2018uky,Lezgi:2021qog,Amrahi:2020jqg,Ebrahim:2020qif}. 

\begin{figure}[ht]
\begin{center}
\includegraphics[width=76 mm]{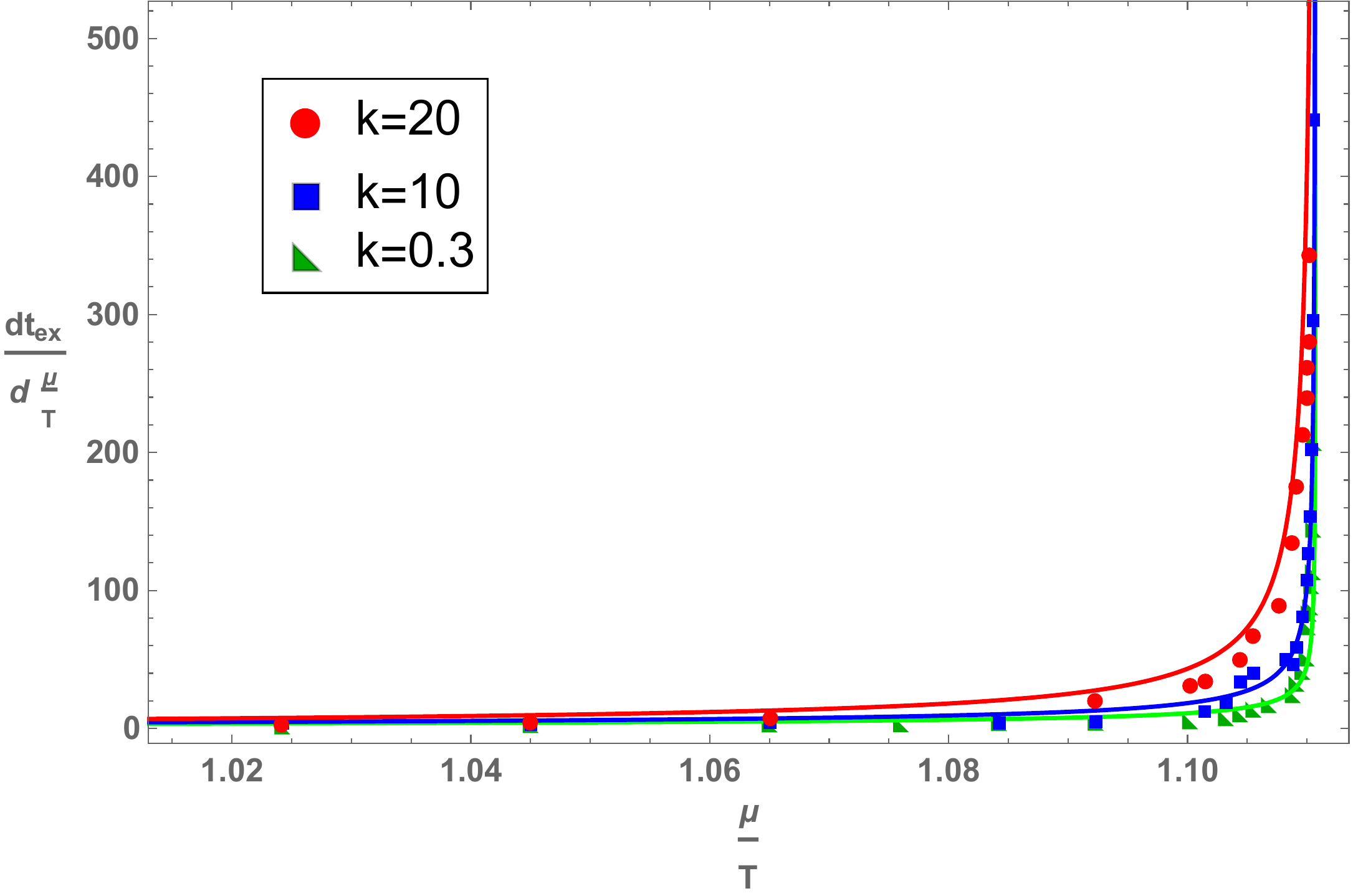}
\caption{ $\frac{dt_{ex}}{d \frac{\mu}{T}}$ as a function of $\frac{\mu}{T}$. We fixed the values of interquark distance $l=0.45$ and the temperature $T=0.31$ ($lT=0.14$) for all cases while for green points the value of the transition time is $k=0.3$  and for blue (red) points are $k=10$ ($k=20$), respectively.
 The green curve is the function, $(\frac{\pi}{2\sqrt{2}}-\frac{\mu}{T})^{-\theta}$, fitted with the data with $\theta=0.529178$. The blue and red curves are the same functions fitted with the data with $\theta=0.641450$ and $\theta=0.830707$, respectively.
\label{f2}}
\end{center}
\end{figure}

To investigate the effect of different values of the transition time on the dynamical critical exponent the behavior of $\frac{dt_{ex}}{d \frac{\mu}{T}}$ as a function of $\frac{\mu}{T}$ is illustrated in figure \ref{f2}. We fixed the values of interquark distance $l=0.45$ and the temperature $T=0.31$ ($lT=0.14$) for all cases while for green points the value of the transition time is $k=0.3$  and for blue (red) points are $k=10$ ($k=20$), respectively. We obtained $\theta=0.529178$ for fast quench $k=0.3$ and $\theta=0.641450$ ($\theta=0.830707$) for slow quenches $k=10$ ($k=20$), respectively. 
 In comparison  with figure \ref{f1}, a considerable change in the value of the dynamical critical exponent $\theta$ is observed in the figure \ref{f2}. The important point is that the dynamical critical exponent $\theta$ is more sensitive to the more larger values of the transition time $k$, that is  the more slower energy injection, the more larger deviations in $\theta$. Therefore, it is seen that just for fast quenches, ($k\ll 1$) the value of the dynamical critical exponent, $\theta$ can be in good agreement with the result that is reported in \cite{Finazzo:2016psx}.
\begin{figure}[ht]
\begin{center}
\includegraphics[width=59 mm]{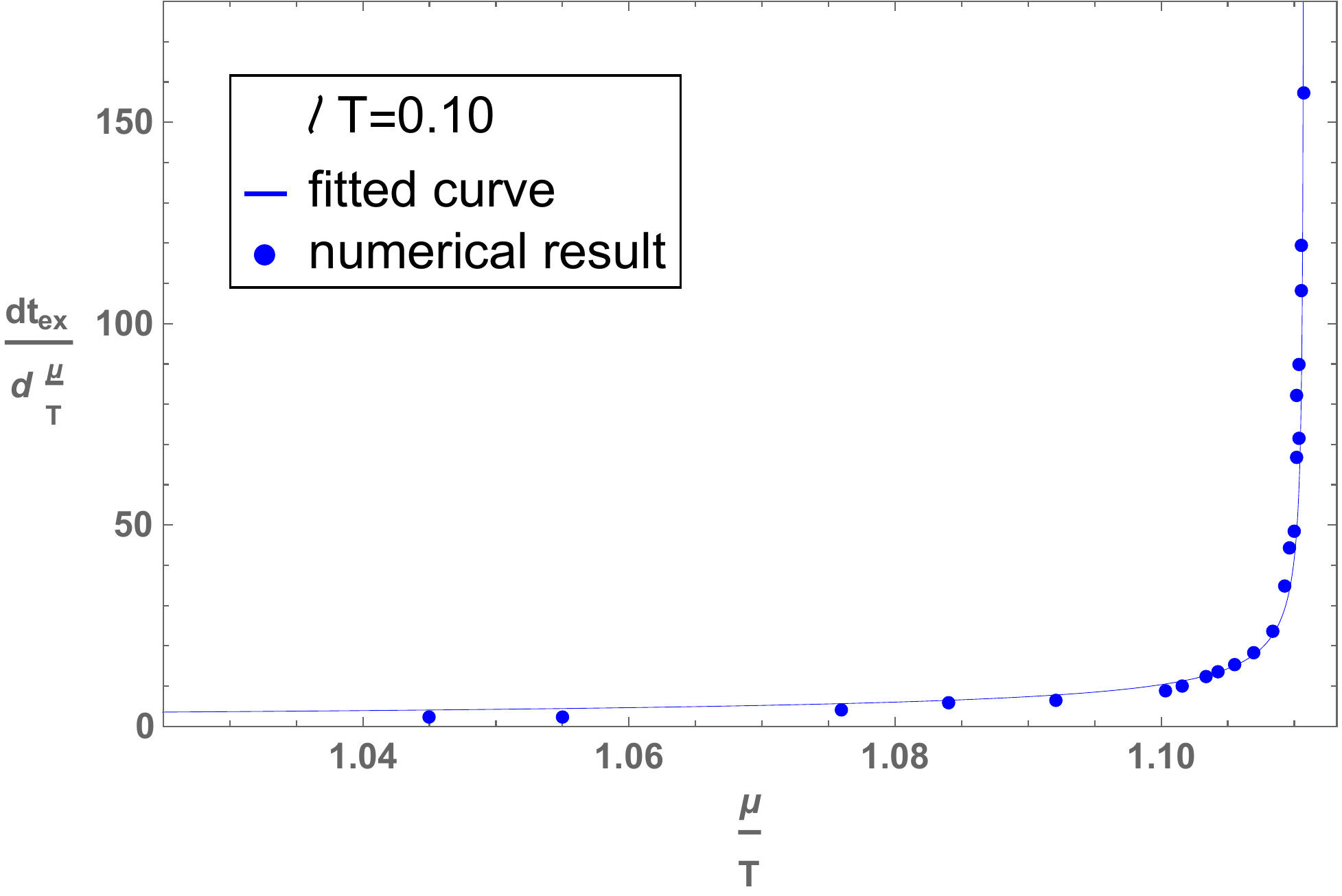}
\includegraphics[width=59 mm]{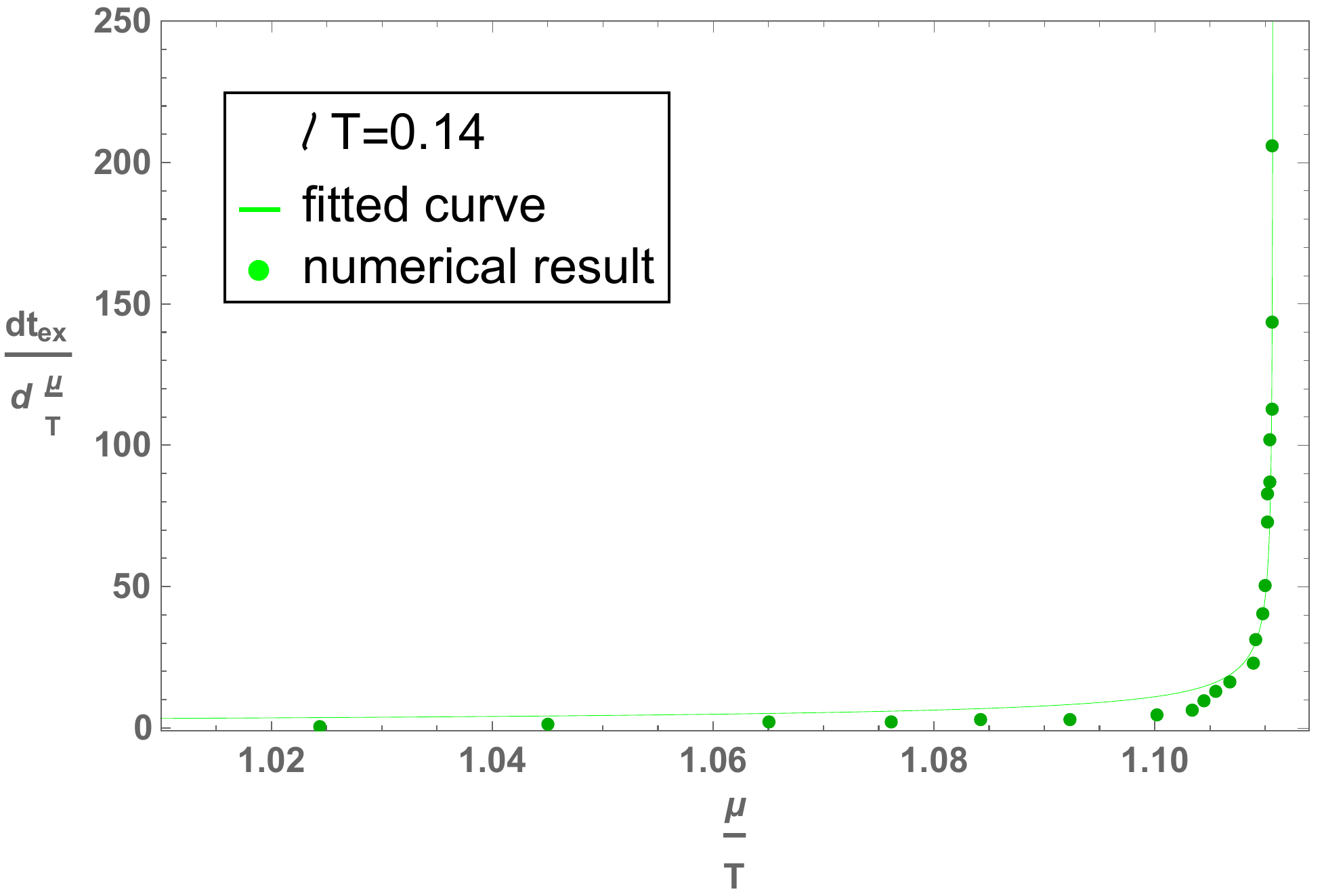}
\includegraphics[width=59 mm]{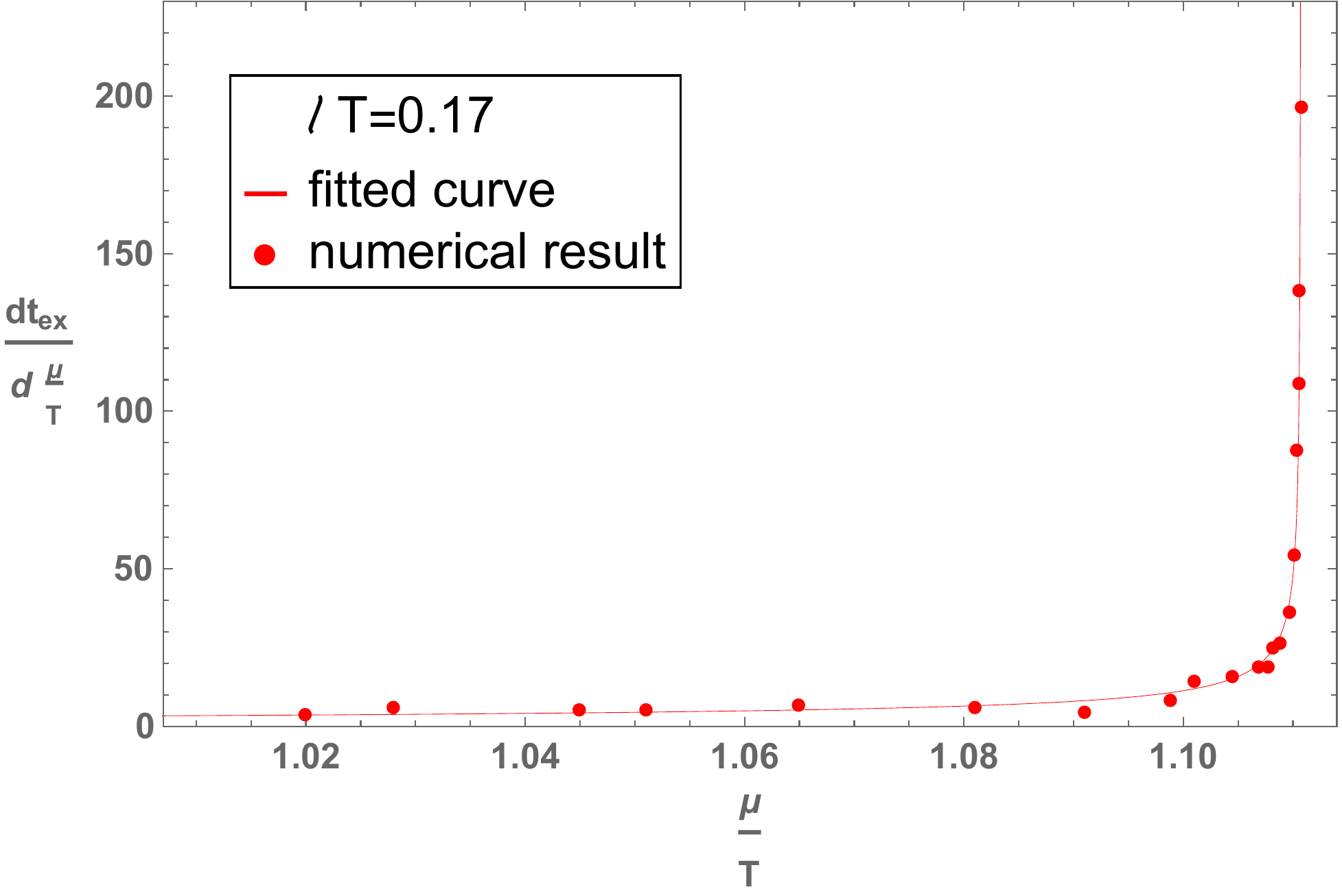}
\caption{   $\frac{dt_{ex}}{d \frac{\mu}{T}}$ as a function of $\frac{\mu}{T}$.
Left: We fixed the values of interquark distance $l=0.27$, the transition time $k=0.3$ , the temperature $T=0.37$  and $lT=0.10$. The blue curve is the function, $(\frac{\pi}{2\sqrt{2}}-\frac{\mu}{T})^{-\theta}$, fitted with the data with $\theta=0.515667$.
Middle: We fixed the values of interquark distance $l=0.45$, the transition time $k=0.3$ , the temperature $T=0.31$  and $lT=0.14$.  The green curve is the same function fitted with the data with $\theta=0.529178$. 
Right: We fixed the values of interquark distance $l=1$, the transition time $k=0.3$ , the temperature $T=0.17$  and $lT=0.17$.  The red curve is the same function fitted with the data with $\theta=0.535088$. 
\label{f3}}
\end{center}
\end{figure}

Now we would like to analyze the effect of interquark distance $l$ on the dynamical critical exponent $\theta$. To do so, the behavior of $\frac{dt_{ex}}{d \frac{\mu}{T}}$ as a function of $\frac{\mu}{T}$ is illustrated in figure \ref{f3} for fixed values of the transition time $k=0.3$. In the left panel we fixed the interquark distance $l=0.27$  and the temperature $T=0.37$ or $lT=0.10$ and obtained the dynamical critical exponent $\theta=0.515667$. 
In the middle panel we fixed the interquark distance $l=0.45$ and the temperature $T=0.31$ or $lT=0.14$ and observed the dynamical critical exponent $\theta=0.529178$. In the right panel we fixed the interquark distance $l=1$   and the temperature $T=0.17$ or $lT=0.17$ and obtained the dynamical critical exponent $\theta=0.535088$. 
The important feature is that by fixing the time interval of energy injection, $k$, there is a low increase in $\theta$ when the value of $lT$ increases.
 In other words, the dynamical critical exponent $\theta$ is sensitive to the value of interquark distance $l$ very smoothly. Therefore, we observed that the value of dynamical critical exponent $\theta$ is more sensible to the value of energy injection $k$ rather than the interquark distance $l$.
 
 We would like to emphasize that for all  numerical results that is obtained in this research we chose the value of $lT$ in such a way that $lT\ll 1$. This is because for the condition $lT\gg 1$ there is no meson bound state and only for the condition $lT\ll 1$
the  meson bound state can exist in the QGP  and hence the meson excitation time, $t_{ex}$, can be defined. 


\section{Discussion and outlook} \label{outloook}

We have continued the research set out in \cite{Hajilou:2018dcb} and utilized the idea of meson excitation time, i.e. $t_{ex}$ to find out some informations from the phase structure of the QCD. It is important to note that, the final results of the paper \cite{Hajilou:2018dcb} has two main parts. In the first part, the effect of the parameters of the theory (such as temperature and chemical potential) on the characteristics of the oscillation, i.e. frequency and amplitude have been investigated. Then, in the second part, after defining the excitation time of the meson, i.e. $t_{ex}$  the effect of various parameters of the theory on the $t_{ex}$ have been described.
But, in this research we borrowed the idea of $t_{ex}$ from \cite{Hajilou:2018dcb} to investigate whether the meson excitation time, $t_{ex}$, can probe the critical point when the system evolves towards the critical point? and what would be the associated dynamical critical exponent? In addition, we study the effect of parameters of the theory and different quenches on the associated dynamical critical exponent and compare our results with other papers.
 Note also that the gravity background in this research is completely different from \cite{Hajilou:2018dcb}. In this work, we consider a particular background that possesses the critical point in such a way that we could investigate the holographic critical point in the field theory side.

In this work we studied the dynamics of a open string attached to the AdS boundary of the Vaidya black hole spacetime in the Einstein-Maxwell dilaton theory. In this theory the black hole solutions possess the "critical point" at which the thermodynamical stability of a black hole solution switches. 
We obtained that the relaxation of the string slows down when the background spacetime is taken to the critical point. We would like to emphasize that, at each value of the ratio of chemical potential and temperature after the injection of energy, relaxation of string slows down and we could calculate the excitation time of meson, $t_{ex}$ to probe the critical point of the theory and then one can obtain the associated dynamical critical exponent.

An important point that we would like to emphasize is that since we work in the probe limit, i.e. the effects of the backreaction of meson are not included, therefore the energy of the meson does not dissipate in the plasma. Consequently, excitation of the meson will remain without decay and the meson's characteristics of oscillation, i.e. frequency and amplitude remain unchanged.
Another point is that, the string dynamics studied in this work relaxed to the equilibrium states. This is very interesting result that has been addressed in \cite{Ali-Akbari:2015ooa,Hajilou:2017sxf,Hajilou:2018dcb}.
Let's consider a bound-state of the stable meson at $t = 0$  which is in its ground state. It is described by the static string hanged from the boundary into the bulk with the end points on the boundary. Due to the injection of energy into the plasma, the temperature and the chemical potential are increased from zero to finite values of $T$ and $\mu$, respectively. According to the results of \cite{Ishii:2014paa,Ali-Akbari:2015ooa}, as the temperature increases the shape of the brane changes time-dependently. As the energy is being injected the turning point of the string gets closer to the black hole horizon. Their results shows that the string oscillates around the string static solution corresponding to the final temperature of the system after finishing the quench. These oscillations are described as the oscillations of the time-dependent Wilson loop in the field theory side. In other words, based on the \cite{Ishii:2014paa} where the authors have studied the dynamics of the shape of the brane in a time-dependent background, these oscillations can be interpreted as if the quench puts the stable meson into the final excited state. The power spectrum of the condensation oscillations gives the excited mesonic modes in the field theory side. As a matter of fact, after the energy injection, the string is oscillating in one of the its normal modes, i.e. relaxes to the equilibrium state or equivalently the quark-antiquark bound state has been excited and oscillates with specific frequency and amplitude.


It is important to note that, we have chosen the stable branch of background solution. This is due to the fact that, the thermal stability conditions are determined by the sign of the heat capacity $C_V$. The positivity of the heat capacity guarantees thermally stable solutions, while the negative heat capacity determines the  unstable solutions. In black hole case, when we decrease $z_h$, i.e. larger horizon, the black hole should be more warmer, while the unstable branch of the background spacetime gets cooler. Therefore, we choose the stable (physical) background solution to investigate the physics of open string as a probe on this background. But, about the solutions of open string in this background  there is one stable and one unstable static solution for the same boundary condition \cite{Hashimoto:2018fkb}. Note also that, the unstable classical configuration is not the minimum energy configuration but is used in \cite{Hashimoto:2018fkb} to probe the horizon. 

In a next phase of research, one interesting direction to extend this work and obtain results of this research from other view point is to consider special initial conditions for open string. In other words, it could be more transparent just to study $t_{ex}$ for the motion of the string with initial (nonlinear) perturbations on a static black hole background, rather than inducing string motion by using the Vaidya (time-dependent) spacetime as the background. This approach originates from the conjecture that the behavior of the relaxation of the string is more or less insensitive to the precise shape of the string excitation from the equilibrium shape.
Another point is that, the critical point corresponds to the background spacetime that is thermodynamically marginally stable. Since, it is the property of the background spacetime, its stability is not directly related to properties of string dynamics on this background spacetime. Therefore, the physical origin of the behaviors found in this work about string dynamics near critical point, could be explained based on the properties of the background geometry and the classical mechanics of the string dynamics in it.


\section*{Acknowledgement}
A. H. would like to thank David Dudal for warm hospitality at KU Leuven Kulak Campus that some parts of this research are accomplished  under a scholarship that was awarded by the Ministry of Science, Research and Technology (Department of Scholarship and Students' s Affairs Abroad) of the Islamic Republic of Iran. Also, thanks to Mohammad Ali-Akbari, Hajar Ebrahim, Irina Aref'eva, Keiju Murata, Song He, Subhash Mahapatra, Seyed Naseh Sajadi, Farid Charmchi, Povel Slepov, and Siddhi Jena for valuable discussions. In addition, thanks to Leila Shahkarami and Farid Charmchi for supporting the Mathematica program to solve equations of motion via the Finite Difference Method.
\appendix
\section{Review on the charged black hole backgrounds} \label{c1}
\subsection{ EM-dilaton black hole background}
Consider the action of 5-dimensional Einstein-Maxwell-dilaton gravity \cite{Zhang:2015dia}:
\be \label{action1}
S=-\frac{1}{16\pi} \int d^{5 }x \sqrt{-g} [ {\cal {R}} - \frac{4}{3}(\nabla\Phi)^2 -V(\Phi)-e^{-\frac{4\alpha}{3}\Phi}\hat{F}_{\mu\nu}\hat{F}^{\mu\nu} ]~.
\ee%
where ${\cal{R}}$ is Ricci scalar, $\hat{F}_{\mu\nu}$ is field strength of the $U(1)$ gauge field, $\Phi$ is the scalar dilaton field, $\alpha$ is the coupling constant between the dilaton and the Maxwell field and $V(\Phi)$ is dilaton potential. For more details see \cite{Zhang:2015dia}. The equations of motion are:
\be\label{metric111}\begin{split}%
 {\cal {R}}_{\mu\nu}-\frac{1}{2}g_{\mu\nu}{\cal {R}}-\frac{4}{3}\left({\partial}_{\mu}\Phi {\partial}_{\nu}\Phi -\frac{1}{2}g_{\mu\nu}(\nabla\Phi)^2-\frac{3}{8}g_{\mu\nu}V(\Phi)\right)-2e^{\frac{-4{\alpha}\Phi}{3}}\left(\hat{F}_{\mu\rho}\hat{F}_{\nu}^{\rho}-\frac{1}{4}g_{\mu\nu}\hat{F}_{\rho{\sigma}}\hat{F}^{\rho{\sigma}}\right)&=0 ~,\\ 
{\nabla}^2{\Phi}-\frac{3}{8}\frac{{\partial V}}{\partial \Phi}+\frac{\alpha}{2}e^{\frac{-4{\alpha}\Phi}{3}}\hat{F}_{\rho{\sigma}}\hat{F}^{\rho{\sigma}}&=0 ~,\\
{\nabla}_{\mu}\left(e^{\frac{-4{\alpha}\Phi}{3}}\hat{F}^{\mu \nu}\right)&=0~.
\end{split}\ee%
 The solution for the metric obtaining from the above action is:
\be \begin{split} \label{metric1}
 ds^2=&-N(z) f(z) dt^2 +\frac{1}{z^4}\frac{dz^2}{(1+b^2z^2)f(z)}+ \frac{1+b^2z^2}{z^2}g(z)d\vec{x}^2 ~,\\ 
f(z)=&\frac{1+b^2z^2}{z^2}\Gamma^{2\gamma}-M\frac{z^2}{1+b^2z^2}\Gamma^{1-\gamma} ~,
\end{split}
\ee%
where $z$ is the radial coordinate in bulk gravity, so that $z=0$ is boundary of the gravity theory where the field theory lives and $\vec{x} \equiv (x_1,x_2,x_3)$. Also, the other functions and parameters are already defined in equation (\ref{metric130}). This solution is asymptotically  AdS$_5$. We set the AdS radius to be one. When we set  $\alpha=0$ in equation (\ref{metric111}), the solution reduces to the well-known Reissner-Nordstr\"{o}m-AdS black hole.

%
\subsection{ EM-dilaton-Vaidya background}
The generalization of the static background (\ref{metric1}) to the time-dependent case can be achieved by adding external source terms to the action (\ref{action1}). To do so, we need introduce external matter sources and then the equations of motion are:
\be\label{metric10}\begin{split}%
 {\cal {R}}_{\mu\nu}-\frac{1}{2}g_{\mu\nu}{\cal {R}}-\frac{4}{3}\left({\partial}_{\mu}\Phi {\partial}_{\nu}\Phi -\frac{1}{2}g_{\mu\nu}(\nabla\Phi)^2-\frac{3}{8}g_{\mu\nu}V(\Phi)\right)-2e^{\frac{-4{\alpha}\Phi}{3}}\left(\hat{F}_{\mu\rho}\hat{F}_{\nu}^{\rho}-\frac{1}{4}g_{\mu\nu}\hat{F}_{\rho{\sigma}}\hat{F}^{\rho{\sigma}}\right)&=8\pi T_{\mu\nu}^{ext},\\ 
{\nabla}_{\mu}\left(e^{\frac{-4{\alpha}\Phi}{3}}\hat{F}^{\mu \nu}\right)&=8\pi J_{ext}^{\nu}~.
\end{split}\ee%
The equation of motion of the dilaton field will not change and are the same as equation (\ref{metric111}). The solution for the EM-dilaton-Vaidya metric in Eddington-Finkelstein coordinates is equation(\ref{metric120}) provided that the external matter source satisfies
\be\label{masscharge}\begin{split}%
8\pi T_{\mu\nu}^{ext} &=\left[\frac{z^3(6+4b^2 z^2)}{4(1+b^2 z^2)^{\frac{3}{2}}}\dot M(\bar{v})\right]{\delta}_{\mu}^{\bar{v}}{\delta}_{\nu}^{\bar{v}} ~,\\
8\pi J_{ext}^{\nu} &= \left[\frac{bz^5}{2}\left(\frac{1}{1+b^2z^2}\right)^{\frac{1}{3}}\frac{\dot M(\bar{v})}{\sqrt{M(\bar{v})}}\right]{\delta}_{z}^{\bar{v}}~,
\end{split}
\ee%
where $\dot{M}(\bar{v})=dM/d\bar{v}$. For more details see \cite{Zhang:2015dia}.

\section{Boundary condition} \label{a1}
The boundary conditions for solving the equations of motion (\ref{eom}) can be obtained by fixing the diffeomorphism invariance on the two-dimensional world-sheet of the string as done in \cite{Ishii:2014paa}. Therefore, on the boundary one can choose $u=v$ for fixing one of the endpoints of string and $u=v+L$ for fixing the other one. Therefore, the boundary condition at the AdS boundary for $Z$ and $X$ are:
\be\begin{split}%
Z|_{u=v}&=0~;~~~~X|_{u=v}=\frac{-l}{2} ~,\cr
Z|_{u=v+L}&=0~;~~~~X|_{u=v+L}=\frac{l}{2} ~.
\end{split}\ee
By applying these boundary conditions, during the injection of energy, the distance between the quark and antiquark will not change. One can get the rest of the boundary conditions by expanding $V(u,v)$, $Z(u,v)$ and $X(u,v)$ about the point $u=v$ at the boundary. As follows
\bea%
V(u,v)&=&V_0(v) + V_1(v) (u-v) + ...~,~~~~~~~~\\ 
Z(u,v)&=&Z_1(v) (u-v) +Z_2(v) (u-v)^2+ ...~,~~~~~~~~\\
X(u,v)&=&\frac{-l}{2} + X_1(v) (u-v) + ...~.~~~~~~~~
\eea%
By putting the above equations into the dynamical equations (\ref{eom}) and demanding the regularity condition at $u=v$, the other boundary conditions one can obtained. The consistency with the constraint equations (\ref{cons}) should be checked. For the other point, $u=v+L$, the mentioned procedure should be followed. Finally, the result of the expansion for the point $u=v$ are:
\bse\begin{align}%
\label{bc1}
V(u,v)&=V_0(v) + {\cal{O}}\left((u-v)^5\right),\\ 
\label{bc2} 
Z(u,v)&=\frac{\dot{V}_0(v)}{2} (u-v) + \frac{\ddot{V}_0(v)}{4} (u-v)^2+\frac{\dddot{V}_0(v)}{12} (u-v)^3 + {\cal{O}}\left((u-v)^4\right),\\ 
X(u,v)&=\frac{-l}{2} + {\cal{O}}\left((u-v)^3\right).
\end{align}\ese%
From the above one can see that
\be%
Z_{,uv}|_{u=v} = 0,\ 2 Z_{,u}|_{u=v} = \dot{V}_0(v) ~,
\ee%
where $\dot{V}(v)=\frac{dV(v)}{dv}$. It is easy to check that the same results for the other point at the boundary, i.e. $u=v+L$ can be obtained. The interested reader can refer to \cite{Ishii:2014paa} for more details. 
\section{Initial condition} \label{b1}
The initial condition for $Z$, $V$ and $X$ can be obtained by using the static equation (\ref{staticz}) and constraint equations (\ref{cons}). In the equation (\ref{staticz}) we replace $x$ and $z$ with the capital ones and set $b=0$, $N(z)=1$, $g(z)=1$ and $f(z)=\frac{1}{z^2}$. Using (\ref{bc1}) , (\ref{bc2}) and considering $V_{,v}>0$ at the boundary, the other conditions $Z_{,u}>0$ and $Z_{,v}<0$ can be found. By imposing $X_{,u}|_{Z=0}=X_{,v}|_{Z=0}=0$ and using the constraint equations (\ref{cons}), we get
\bea%
\label{eqV1}
V_{,u}
&= Z_{,u} \bigg(-1+\sqrt{1+\big(\frac{dX}{dZ}\big)^2}\bigg) ~,~\\
\label{eqV2}
V_{,v}
&= Z_{,v} \bigg(-1-\sqrt{1+\big(\frac{dX}{dZ}\big)^2}\bigg) ~.
\eea%
Now by taking the derivative of equation (\ref{eqV1}) with respect to $v$ and equation (\ref{eqV2}) with respect to $u$ and considering that $V_{,uv}=V_{,vu}$ the result is
\be\label{equ}%
\bigg( Z_{,u} \sqrt{1+\big(\frac{dX}{dZ}\big)^2} \bigg)_{,v}=0 ~.
\ee%
Using equation (\ref{staticz}) and substituting $\frac{dX}{dZ}$ into the above equation one can get the initial condition for $Z(u,v)$
\be%
\label{solZ}
Z \, _2F_1\bigg(\frac{1}{2},\frac{1}{4};\frac{5}{4};\frac{Z^{4}}{{Z_*}^{4}}\bigg) = \phi(u) - \phi(v) ~,
\ee%
where $\phi(y)$ is an arbitrary function and our choice is $\phi(y)=y$ \cite{Ishii:2014paa}.
One can get the initial condition for $X(u,v)$ by integrating the equation (\ref{staticz}) as follows
\be%
X(u,v) = \frac{l}{2} - \frac{Z^3}{3 {Z_*}^{2}} \, _2F_1\left(\frac{1}{2},\frac{3}{4};\frac{7}{4};\frac{Z^{4}}{Z_*^{4}}\right) ~.
\ee%
Where, $Z_*$ is the turning point of the string and since at $Z=Z_*$ the equation $X(u,v)=0$, we have
\be
Z_*=\frac{3~l}{2~ _2F_1\left(\frac{1}{2},\frac{3}{4};\frac{7}{4};1\right)}~.
\ee
Also, using the equations (\ref{eqV1}) and (\ref{eqV2}) one can obtain the initial condition for $V(u,v)$:
\bea%
V(u,v)&=- Z \left(1-\, _2F_1\left(\frac{1}{2},\frac{1}{4};\frac{5}{4};\frac{Z^{4}}{Z_*^{4}}\right)\right)+\chi(v) ~,~~~~~~~\\
V(u,v)&=- Z \left(1+\, _2F_1\left(\frac{1}{2},\frac{1}{4};\frac{5}{4};\frac{Z^{4}}{Z_*^{4}}\right)\right)+{\tilde{\chi}}(u) ~,~~~~~~~
\eea%
where $\chi$ and $\tilde{\chi}$ are arbitrary functions. Moreover, by equalizing the above equations and using (\ref{solZ}), we would have
\be%
\chi(v) = 2 \phi(v) ~,~~~~~{\tilde{\chi}}(u) = 2 \phi(u) ~.
\ee%
For more details see \cite{Ishii:2014paa}.
\section{Numerical procedure to calculate the critical exponent} \label{numeric}
In this section we discuss the numerical procedure that is used above to determine the dynamical critical exponent  $\theta$ when the physical system moves towards the critical point  $(\frac{\mu}{T})_{\star}=1.11072$. At first, for producing the $t_{ex}$ for each point, i.e. the finite value of $\frac{\mu}{T}$, we used equation (\ref{fit11}). In addition, to calculate the dynamical critical exponent $\theta$ let's take an specific example which the parameters of our physical system are as the transition time $k=0.3$, interquark distance $l=0.45$ and $l T=0.14$. Therefore, when our system is moving towards the critical point from point $\frac{\mu}{T}=1.0243$ to $(\frac{\mu}{T})_{\star}=1.11072$ we split this interval into $22$ subintervals with different step sizes ($\Delta (\frac{\mu}{T})$ that are between $10^{-2}$ and $10^{-5}$). Also, for computing the numerical derivatives we used the slope (\ref{cc}). In Table \ref{tab:table2} we specified the subintervals utilized to determine the dynamical critical exponent $\theta$. 
 The result corresponding with above fixed parameters is shown in the middle panel of the figure  \ref{f3}  and (after using Wolfram's Mathematica) the associated critical exponent $\theta=0.529178$ is obtained.
\begin{table}[h!]
  \begin{center}
    \begin{tabular}{l|c|c|r} 
      \textbf{~~~point~~~} & \textbf{~~~~Value of $t_{ex}$~~~~}& \textbf{~~~~Value of $\frac{\mu}{T}$~~~~} & \textbf{~~~~Value of $\frac{d t_{ex}}{d\frac{\mu}{T}}$~~~~}\\
      \hline
     ~~~~~~~ 1 & 0.7101& 1.0243& 0.64~~~~~~~~~~\\
      ~~~~~~~ 2 & 0.7234& 1.0451& 1.64~~~~~~~~~~\\
      ~~~~~~~ 3 & 0.7561& 1.0651 & 2.12~~~~~~~~~~\\
      ~~~~~~~ 4 & 0.7794& 1.0761 & 2.20~~~~~~~~~~\\
       ~~~~~~~ 5 & 0.7974 & 1.0843 & 3.10~~~~~~~~~~\\
      ~~~~~~~ 6 & 0.8225 &1.0924 & 3.34~~~~~~~~~~\\
       ~~~~~~~ 7 & 0.8489 &1.1003 & 4.40~~~~~~~~~~\\
      ~~~~~~~ 8 & 0.8621 &1.1033 & 6.67~~~~~~~~~~\\
      ~~~~~~~ 9 & 0.8701 &1.1045 & 9.45~~~~~~~~~~\\
      ~~~~~~~10 & 0.8805 &1.1056 & 12.85~~~~~~~~~~\\
       ~~~~~~~11& 0.8972 &1.1069 & 16.15~~~~~~~~~~\\
      ~~~~~~~12& 0.9295 &1.1089 & 23.33~~~~~~~~~~\\ 
      ~~~~~~~13& 0.9365 &1.1092 &31.60~~~~~~~~~~\\
      ~~~~~~~14& 0.9523 &1.1097 & 40.29~~~~~~~~~~\\
      ~~~~~~~15& 0.9664 &1.11005 & 50.00~~~~~~~~~~\\
       ~~~~~~~16& 0.9714 &1.11015 & 72.50~~~~~~~~~~\\
      ~~~~~~~17& 0.9772 &1.11023 & 82.50~~~~~~~~~~\\
      ~~~~~~~18& 0.9871 &1.11035 &87.00~~~~~~~~~~\\
      ~~~~~~~19& 0.9958 &1.11045 & 102.31~~~~~~~~~~\\
      ~~~~~~~20& 1.0091 &1.11058 & 112.50~~~~~~~~~~\\
      ~~~~~~~21& 1.0136 &1.11062 & 143.28~~~~~~~~~~\\
        ~~~~~~~22& 1.0232 &1.11069 & 206.06~~~~~~~~~~\\
         ~~~~~~~23& 1.0300 &1.11072 &     ~~~~~~~~~~\\
    \end{tabular}
        \caption{Values of $t_{ex}$,  $\frac{\mu}{T}$ and $\frac{d t_{ex}}{d\frac{\mu}{T}}$ used for the fit procedure for the transition time $k=0.3$, interquark distance $l=0.45$ and $l T=0.14$ with the fitted critical exponent $\theta=0.529178$.}
    \label{tab:table2}
  \end{center}
\end{table}
\section{Meson potential at zero temperature} \label{zero17}
\begin{figure}[ht]
\begin{center}
\includegraphics[width=72 mm]{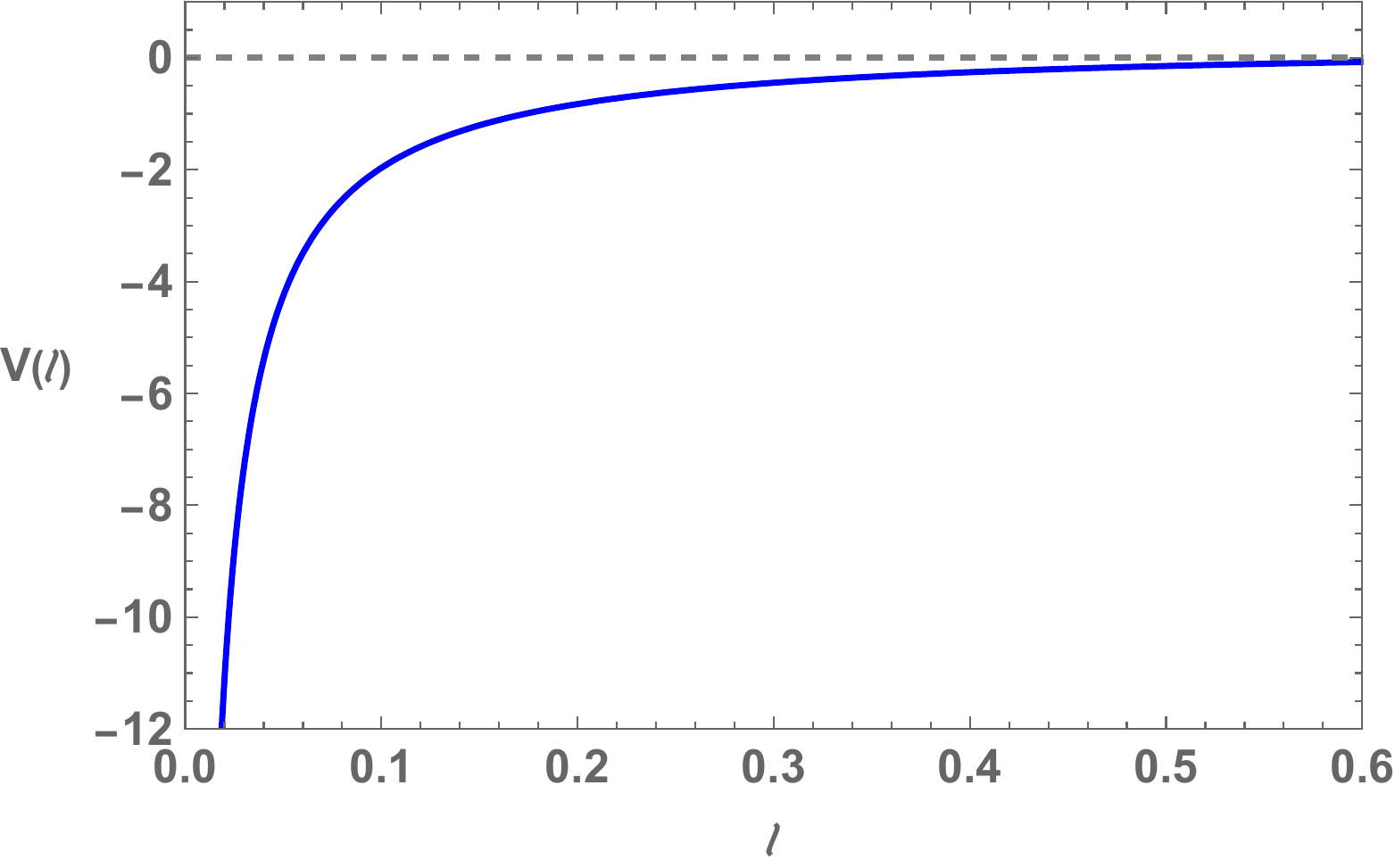}
\caption{  Potential of meson in the $AdS_5$ background that corresponds with zero temperature in field theory. Utilizing the Wilson loop the Coloumb potential can be obtained.
\label{free33}}
\end{center}
\end{figure}

In this research, before the injection of energy the spacetime background is pure AdS$_5$. Utilizing the equations (\ref{wilson3}), (\ref{wilson4}) and (\ref{action2}) one can find the static potential of meson in the AdS$_5$ background that is correspondence with zero temperature field theory. After some algebra one can find:
\begin{align}\label{finallstatic}
V(l) = \frac{1}{\pi\alpha^\prime} \bigg[\frac{1}{z_\ast}& \int_1^\infty\,dy 
\left(\sqrt{\frac{y^{4}-y_h^{4}}{y^{4}-1}}-1\right) 
-\left(\frac{1}{z_{\ast}}-\frac{1}{z_h}\right)\bigg]~,
\end{align}
where, $y=z/z_h$ and $y_h=z_\star/z_h$. The static potential of the meson at zero temperature is depicted in figure \ref{free33}.
By considering the Cornell potential \cite{Eichten:1978tg,Andreev:2006ct,Bruni:2018dqm,Yang:2015aia}:
\be \label{cornel}
V(l)=-\frac{\kappa}{l}+ \sigma_s l + C ~,
\ee%
where, $\kappa$ is a Coulomb strength parameter, $\sigma_s$ is QCD string tension and $C$ is a constant. 
We observe that here at zero temperature, since the spacetime is pure AdS$_5$ then surely when we calculate the potential of meson via Wilson loop, we will obtain just Coulomb potential part ($-\frac{\kappa}{l}$) of the Cornell Potential. 
This is because that the linear regime of the Cornell potential can be obtained when the geometry is ended in the holographic direction $z$ \cite{Witten:1998zw,Polchinski:2001tt,CasalderreySolana:2011us}. But, in the case of pure AdS$_5$ the spacetime does not end in the direction of $z$ in such a way that Wilson loop potential does not possess the linear regime of the Cornell potential.

\section{Time evolution of the string} \label{stringplot22}
\begin{figure}[ht]
\begin{center}
\includegraphics[width=98 mm]{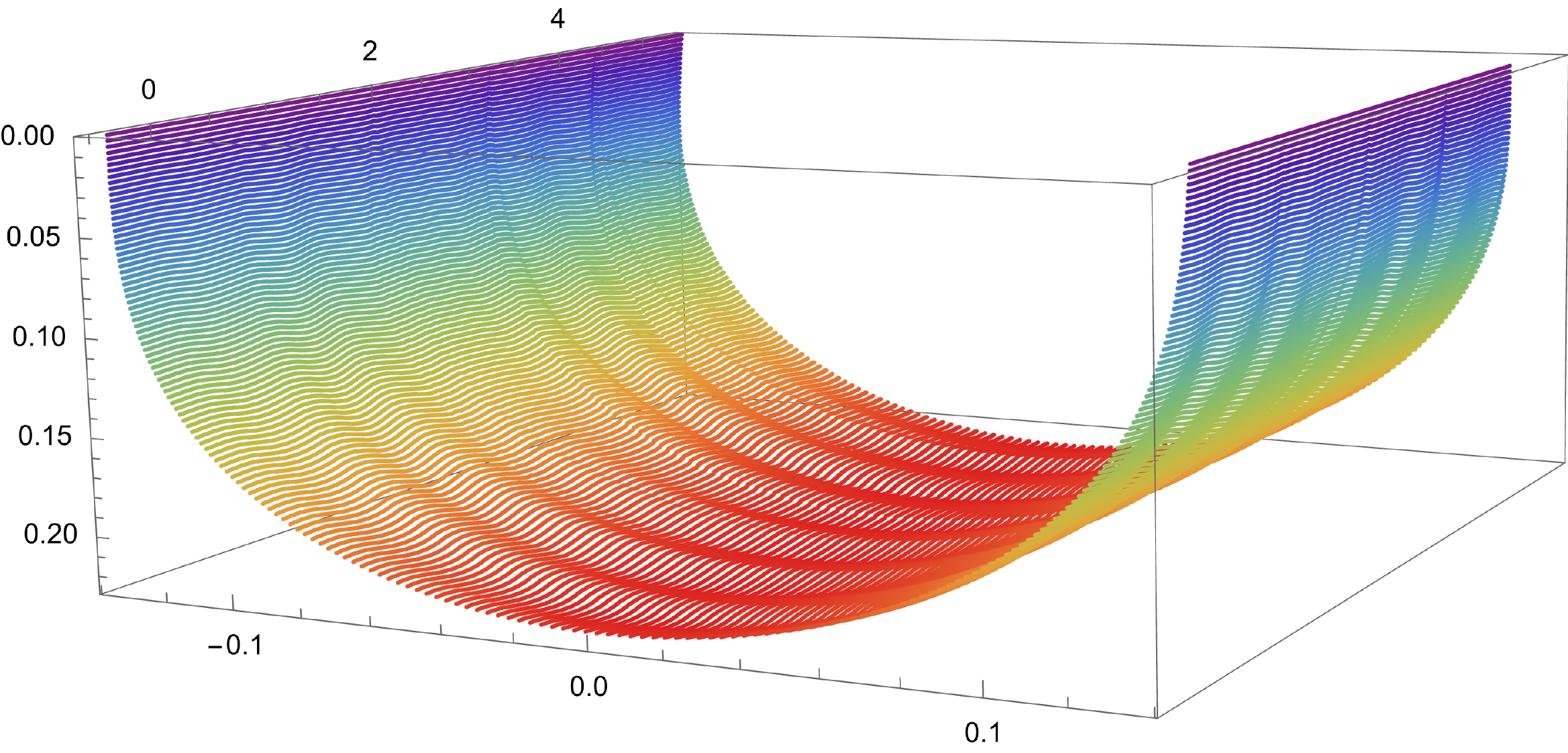}
\caption{ Time evolution of the classical open string in the EM-dilaton-Vaidya background. We have fixed the interquark distance $l=0.27$, transition time $k=0.3$ and the temperature $T=0.37~ (lT=0.10)$. Here, we have $\frac{\mu}{T}=1.11071$
\label{string3d}}
\end{center}
\end{figure}

We depicted the time evolution of the classical open string in the EM-dilaton-Vaidya background, in figure (\ref{string3d}) for fixed values of the interquark distance $l=0.27$, transition time $k=0.3$ and the temperature $T=0.37~ (lT=0.10)$. As shown in  in figure (\ref{string3d}), in the gravity side we see that the open string oscillates around the static configuration correspondence with the static potential of the meson after injection of energy in the field theory side. In the figure (\ref{string3d}), we have chosen the $\frac{\mu}{T}=1.11071$, but it is important to note that, this behavior, i.e oscillation around static configuration is observed for all values of the ratio of $\frac{\mu}{T}$  as well as critical point. For each values of $\frac{\mu}{T}$ we can obtain the excitation time of the meson and then the dynamical critical exponent $\theta$ can be calculated.



\begin{thebibliography}{99}
\bibitem{CasalderreySolana:2011us} 
  J.~Casalderrey-Solana, H.~Liu, D.~Mateos, K.~Rajagopal and U.~A.~Wiedemann,
  ``Gauge/String Duality, Hot QCD and Heavy Ion Collisions,''
  book:Gauge/String Duality, Hot QCD and Heavy Ion Collisions. Cambridge, UK: Cambridge University Press, 2014
  [arXiv:1101.0618 [hep-th]].
  
\bibitem{Ammon:2015wua}
M.~Ammon and J.~Erdmenger,
``Gauge/gravity duality: Foundations and applications,''
Cambridge Univ. Pr., Cambridge, UK, 2015.

\bibitem{Shuryak:2003xe} 
  E.~Shuryak,
  ``Why does the quark gluon plasma at RHIC behave as a nearly ideal fluid?,''
  Prog.\ Part.\ Nucl.\ Phys.\  {\bf 53}, 273 (2004)
  [hep-ph/0312227].
  
\bibitem{Shuryak:2004cy} 
  E.~V.~Shuryak,
  ``What RHIC experiments and theory tell us about properties of quark-gluon plasma?,''
  Nucl.\ Phys.\ A {\bf 750}, 64 (2005)
  [hep-ph/0405066].
  
\bibitem{Chesler:2013lia}
P.~M.~Chesler and L.~G.~Yaffe,
``Numerical solution of gravitational dynamics in asymptotically anti-de Sitter spacetimes,''
JHEP \textbf{07}, 086 (2014)
[arXiv:1309.1439 [hep-th]].
  
\bibitem{Bhattacharyya:2009uu}
S.~Bhattacharyya and S.~Minwalla,
``Weak Field Black Hole Formation in Asymptotically AdS Spacetimes,''
JHEP \textbf{09}, 034 (2009)
[arXiv:0904.0464 [hep-th]].  
  
\bibitem{Balasubramanian:2010ce}
V.~Balasubramanian, A.~Bernamonti, J.~de Boer, N.~Copland, B.~Craps, E.~Keski-Vakkuri, B.~Muller, A.~Schafer, M.~Shigemori and W.~Staessens,
``Thermalization of Strongly Coupled Field Theories,''
Phys. Rev. Lett. \textbf{106}, 191601 (2011)
[arXiv:1012.4753 [hep-th]].


\bibitem{Balasubramanian:2011ur}
V.~Balasubramanian, A.~Bernamonti, J.~de Boer, N.~Copland, B.~Craps, E.~Keski-Vakkuri, B.~Muller, A.~Schafer, M.~Shigemori and W.~Staessens,
``Holographic Thermalization,''
Phys. Rev. D \textbf{84}, 026010 (2011)
[arXiv:1103.2683 [hep-th]].
  
  
\bibitem{Buchel:2014gta}
A.~Buchel, R.~C.~Myers and A.~van Niekerk,
``Nonlocal probes of thermalization in holographic quenches with spectral methods,''
JHEP \textbf{02}, 017 (2015)
[erratum: JHEP \textbf{07}, 137 (2015)]
[arXiv:1410.6201 [hep-th]].
 
 
\bibitem{Rothkopf:2011db}
A.~Rothkopf, T.~Hatsuda and S.~Sasaki,
``Complex Heavy-Quark Potential at Finite Temperature from Lattice QCD,''
Phys. Rev. Lett. \textbf{108}, 162001 (2012)
[arXiv:1108.1579 [hep-lat]].

\bibitem{Maldacena:1997re} 
  J.~M.~Maldacena,
  ``The Large N limit of superconformal field theories and supergravity,''
  Int.\ J.\ Theor.\ Phys.\  {\bf 38}, 1113 (1999)
  [Adv.\ Theor.\ Math.\ Phys.\  {\bf 2}, 231 (1998)]
  [hep-th/9711200].
  
\bibitem{Witten:1998qj} 
  E.~Witten,
  ``Anti-de Sitter space and holography,''
  Adv.\ Theor.\ Math.\ Phys.\  {\bf 2}, 253 (1998)
  [hep-th/9802150].
  
\bibitem{Gubser:1998bc} 
  S.~S.~Gubser, I.~R.~Klebanov and A.~M.~Polyakov,
  ``Gauge theory correlators from noncritical string theory,''
  Phys.\ Lett.\ B {\bf 428}, 105 (1998)
  [hep-th/9802109].
 
\bibitem{Gursoy:2007cb}
U.~Gursoy and E.~Kiritsis,
``Exploring improved holographic theories for QCD: Part I,''
JHEP \textbf{02}, 032 (2008)
[arXiv:0707.1324 [hep-th]].

\bibitem{Gursoy:2007er}
U.~Gursoy, E.~Kiritsis and F.~Nitti,
``Exploring improved holographic theories for QCD: Part II,''
JHEP \textbf{02}, 019 (2008)
[arXiv:0707.1349 [hep-th]].

\bibitem{Dudal:2018ztm}
D.~Dudal and S.~Mahapatra,
``Interplay between the holographic QCD phase diagram and entanglement entropy,''
JHEP \textbf{07}, 120 (2018)
[arXiv:1805.02938 [hep-th]].

\bibitem{Maldacena:1998im} 
  J.~M.~Maldacena,
  ``Wilson loops in large N field theories,''
  Phys.\ Rev.\ Lett.\  {\bf 80}, 4859 (1998)
  [hep-th/9803002].
  
\bibitem{Brandhuber:1998bs} 
A.~Brandhuber, N.~Itzhaki, J.~Sonnenschein and S.~Yankielowicz,
``Wilson loops in the large N limit at finite temperature,''
Phys.\ Lett.\ B {\bf 434}, 36 (1998)
[hep-th/9803137].


\bibitem{Rey:1998bq} 
S.~J.~Rey, S.~Theisen and J.~T.~Yee,
``Wilson-Polyakov loop at finite temperature in large N gauge theory and anti-de Sitter supergravity,''
Nucl.\ Phys.\ B {\bf 527}, 171 (1998)
[hep-th/9803135];
\bibitem{Sonnenschein:2000qm} 
J.~Sonnenschein,
``Stringy confining Wilson loops,''
PoS tmr {\bf 2000}, 008 (2000)
[hep-th/0009146].
\bibitem{Finazzo:2013aoa} 
S.~I.~Finazzo and J.~Noronha,
``Estimates for the Thermal Width of Heavy Quarkonia in Strongly Coupled Plasmas from Holography,''
JHEP {\bf 1311}, 042 (2013)
[arXiv:1306.2613 [hep-ph]].


\bibitem{Ali-Akbari:2015ooa}
M.~Ali-Akbari, F.~Charmchi, A.~Davody, H.~Ebrahim and L.~Shahkarami,
``Evolution of Wilson loop in time-dependent N=4 super Yang-Mills plasma,''
Phys. Rev. D \textbf{93}, no.8, 086005 (2016)
[arXiv:1510.00212 [hep-th]].

\bibitem{Dudal:2017max}
D.~Dudal and S.~Mahapatra,
``Thermal entropy of a quark-antiquark pair above and below deconfinement from a dynamical holographic QCD model,''
Phys. Rev. D \textbf{96}, no.12, 126010 (2017)
[arXiv:1708.06995 [hep-th]].

\bibitem{Cai:2012xh}
R.~G.~Cai, S.~He and D.~Li,
``A hQCD model and its phase diagram in Einstein-Maxwell-Dilaton system,''
JHEP \textbf{03}, 033 (2012)
[arXiv:1201.0820 [hep-th]].


\bibitem{Bohra:2019ebj}
H.~Bohra, D.~Dudal, A.~Hajilou and S.~Mahapatra,
``Anisotropic string tensions and inversely magnetic catalyzed deconfinement from a dynamical AdS/QCD model,''
Phys. Lett. B \textbf{801}, 135184 (2020)
[arXiv:1907.01852 [hep-th]].

\bibitem{Asadi:2021nbd}
M.~Asadi and A.~Hajilou,
``Meson potential energy in a non-conformal holographic model,''
Nucl. Phys. B \textbf{979}, 115744 (2022)
[arXiv:2112.04209 [hep-th]].

\bibitem{Liu:2006ug}
H.~Liu, K.~Rajagopal and U.~A.~Wiedemann,
``Calculating the jet quenching parameter from AdS/CFT,''
Phys. Rev. Lett. \textbf{97}, 182301 (2006)
[arXiv:hep-ph/0605178 [hep-ph]].

\bibitem{Cai:2012eh}
R.~G.~Cai, S.~Chakrabortty, S.~He and L.~Li,
``Some aspects of QGP phase in a hQCD model,''
JHEP \textbf{02}, 068 (2013)
[arXiv:1209.4512 [hep-th]].


\bibitem{Galante:2012pv}
D.~Galante and M.~Schvellinger,
``Thermalization with a chemical potential from AdS spaces,''
JHEP \textbf{07}, 096 (2012)
[arXiv:1205.1548 [hep-th]].


\bibitem{Ali-Akbari:2012gku}
M.~Ali-Akbari and H.~Ebrahim,
``Thermalization in External Magnetic Field,''
JHEP \textbf{03}, 045 (2013)
[arXiv:1211.1637 [hep-th]].

\bibitem{Ali-Akbari:2012fzl}
M.~Ali-Akbari and H.~Ebrahim,
``Meson Thermalization in Various Dimensions,''
JHEP \textbf{04}, 145 (2012)
[arXiv:1203.3425 [hep-th]].

\bibitem{Dey:2015poa}
A.~Dey, S.~Mahapatra and T.~Sarkar,
``Holographic Thermalization with Weyl Corrections,''
JHEP \textbf{01}, 088 (2016)
[arXiv:1510.00232 [hep-th]].


\bibitem{Ali-Akbari:2015bha}
M.~Ali-Akbari, F.~Charmchi, A.~Davody, H.~Ebrahim and L.~Shahkarami,
``Time-dependent meson melting in an external magnetic field,''
Phys. Rev. D \textbf{91}, 106008 (2015)
[arXiv:1503.04439 [hep-th]].

\bibitem{Bohra:2020qom}
H.~Bohra, D.~Dudal, A.~Hajilou and S.~Mahapatra,
``Chiral transition in the probe approximation from an Einstein-Maxwell-dilaton gravity model,''
Phys. Rev. D \textbf{103}, no.8, 086021 (2021)
[arXiv:2010.04578 [hep-th]].

   \bibitem{Ali-Akbari:2013txa}
M.~Ali-Akbari and H.~Ebrahim,
``Chiral symmetry breaking: To probe anisotropy and magnetic field in quark-gluon plasma,''
Phys. Rev. D \textbf{89}, no.6, 065029 (2014)
[arXiv:1309.4715 [hep-th]].

\bibitem{Arefeva:2020vae}
I.~Y.~Aref'eva, K.~Rannu and P.~Slepov,
``Holographic model for heavy quarks in anisotropic hot dense QGP with external magnetic field,''
JHEP \textbf{07}, 161 (2021)
[arXiv:2011.07023 [hep-th]].


\bibitem{Zhou:2020ssi}
J.~Zhou, X.~Chen, Y.~Q.~Zhao and J.~Ping,
``Thermodynamics of heavy quarkonium in a magnetic field background,''
Phys. Rev. D \textbf{102}, no.8, 086020 (2020)
[arXiv:2006.09062 [hep-ph]].


\bibitem{Dudal:2021jav}
D.~Dudal, A.~Hajilou and S.~Mahapatra,
``A quenched 2-flavour Einstein\textendash{}Maxwell\textendash{}Dilaton gauge-gravity model,''
Eur. Phys. J. A \textbf{57}, no.4, 142 (2021)
[arXiv:2103.01185 [hep-th]].


\bibitem{Arefeva:2021mag}
I.~Y.~Aref'eva, K.~Rannu and P.~S.~Slepov,
``Anisotropic solutions for a holographic heavy-quark model with an external magnetic field,''
Teor. Mat. Fiz. \textbf{207}, no.1, 44-57 (2021).

\bibitem{Fang:2021ucy}
Z.~Fang, Y.~Y.~Li and Y.~L.~Wu,
``QCD phase diagram with a background magnetic field in an improved soft-wall AdS/QCD model,''
Eur. Phys. J. C \textbf{81}, no.6, 545 (2021).

\bibitem{Dudal:2016joz}
D.~Dudal and S.~Mahapatra,
``Confining gauge theories and holographic entanglement entropy with a magnetic field,''
JHEP \textbf{04}, 031 (2017)
[arXiv:1612.06248 [hep-th]].

\bibitem{He:2020fdi}
S.~He, Y.~Yang and P.~H.~Yuan,
``Analytic Study of Magnetic Catalysis in Holographic QCD,''
[arXiv:2004.01965 [hep-th]].
\bibitem{Qin:2010nq}
S.~x.~Qin, L.~Chang, H.~Chen, Y.~x.~Liu and C.~D.~Roberts,
``Phase diagram and critical endpoint for strongly-interacting quarks,''
Phys. Rev. Lett. \textbf{106}, 172301 (2011)
[arXiv:1011.2876 [nucl-th]].

\bibitem{DeWolfe:2010he} 
  O.~DeWolfe, S.~S.~Gubser and C.~Rosen,
  ``A holographic critical point,''
  Phys.\ Rev.\ D {\bf 83}, 086005 (2011)
  [arXiv:1012.1864 [hep-th]].
   
 \bibitem{Ebrahim:2017gvk} 
  H.~Ebrahim and M.~Ali-Akbari,
  ``Dynamically probing strongly-coupled field theories with critical point,''
  Phys.\ Lett.\ B {\bf 783}, 43 (2018)
  [arXiv:1712.08777 [hep-th]].

\bibitem{Cai:2022omk}
R.~G.~Cai, S.~He, L.~Li and Y.~X.~Wang,
``Probing QCD critical point and induced gravitational wave by black hole physics,''
Phys. Rev. D \textbf{106}, no.12, L121902 (2022)
[arXiv:2201.02004 [hep-th]].
   
   \bibitem{Zhang:2015dia}
S.~J.~Zhang and E.~Abdalla,
``Holographic Thermalization in Charged Dilaton Anti-de Sitter Spacetime,''
Nucl. Phys. B \textbf{896}, 569-586 (2015)
[arXiv:1503.07700 [hep-th]].
   
\bibitem{Hajilou:2018dcb} 
  A.~Hajilou and M.~Ali-Akbari,
  ``Meson Excitation at Finite Chemical Potential,''
  Eur.\ Phys.\ J.\ C {\bf 79}, no. 3, 254 (2019)
  [arXiv:1804.07965 [hep-th]].

   
\bibitem{DeWolfe:2011ts}
O.~DeWolfe, S.~S.~Gubser and C.~Rosen,
``Dynamic critical phenomena at a holographic critical point,''
Phys. Rev. D \textbf{84}, 126014 (2011)
[arXiv:1108.2029 [hep-th]].

\bibitem{Finazzo:2016psx} 
  S.~I.~Finazzo, R.~Rougemont, M.~Zaniboni, R.~Critelli and J.~Noronha,
  ``Critical behavior of non-hydrodynamic quasinormal modes in a strongly coupled plasma,''
  JHEP {\bf 1701}, 137 (2017)
  [arXiv:1610.01519 [hep-th]].  
  
   
\bibitem{Amiri-Sharifi:2016uso}
S.~Amiri-Sharifi, M.~Ali-Akbari, A.~Kishani-Farahani and N.~Shafie,
``Double Relaxation via AdS/CFT,''
Nucl. Phys. B \textbf{909}, 778-795 (2016)
[arXiv:1601.04281 [hep-th]].

\bibitem{He:2013qq}
S.~He, S.~Y.~Wu, Y.~Yang and P.~H.~Yuan,
``Phase Structure in a Dynamical Soft-Wall Holographic QCD Model,''
JHEP \textbf{04}, 093 (2013)
[arXiv:1301.0385 [hep-th]].

\bibitem{Gursoy:2008za}
U.~Gursoy, E.~Kiritsis, L.~Mazzanti and F.~Nitti,
``Holography and Thermodynamics of 5D Dilaton-gravity,''
JHEP \textbf{05}, 033 (2009)
[arXiv:0812.0792 [hep-th]].

\bibitem{Arefeva:2018hyo}
I.~Aref'eva and K.~Rannu,
``Holographic Anisotropic Background with Confinement-Deconfinement Phase Transition,''
JHEP \textbf{05}, 206 (2018)
[arXiv:1802.05652 [hep-th]].


\bibitem{Gursoy:2018ydr}
U.~G\"ursoy, M.~J\"arvinen, G.~Nijs and J.~F.~Pedraza,
``Inverse Anisotropic Catalysis in Holographic QCD,''
JHEP \textbf{04}, 071 (2019)
[erratum: JHEP \textbf{09}, 059 (2020)]
[arXiv:1811.11724 [hep-th]].

\bibitem{Hawking:1982dh}
S.~W.~Hawking and D.~N.~Page,
``Thermodynamics of Black Holes in anti-De Sitter Space,''
Commun. Math. Phys. \textbf{87}, 577 (1983)

\bibitem{He:2022amv}
S.~He, L.~Li, Z.~Li and S.~J.~Wang,
``Gravitational Waves and Primordial Black Hole Productions from Gluodynamics,''
[arXiv:2210.14094 [hep-ph]].

\bibitem{Li:2017tdz}
M.~W.~Li, Y.~Yang and P.~H.~Yuan,
``Approaching Confinement Structure for Light Quarks in a Holographic Soft Wall QCD Model,''
Phys. Rev. D \textbf{96}, no.6, 066013 (2017)
[arXiv:1703.09184 [hep-th]].

\bibitem{Sajadi:2023zke}
S.~N.~Sajadi,
``Holographic anisotropic background in 5D Einstien\textendash{}Gauss\textendash{}Bonnet gravity,''
Eur. Phys. J. C \textbf{83}, no.1, 89 (2023)
[arXiv:2301.00638 [hep-th]].









\bibitem{Hajilou:2017sxf} 
  A.~Hajilou, M.~Ali-Akbari and F.~Charmchi,
  ``A Classical String in Lifshitz–Vaidya Geometry,''
  Eur.\ Phys.\ J.\ C {\bf 78}, no. 5, 424 (2018)
  [arXiv:1707.00967 [hep-th]].
 

\bibitem{Ishii:2014paa}
T.~Ishii, S.~Kinoshita, K.~Murata and N.~Tanahashi,
``Dynamical Meson Melting in Holography,''
JHEP \textbf{04}, 099 (2014)
[arXiv:1401.5106 [hep-th]].
  
\bibitem{Ageev:2016gtl}
D.~S.~Ageev, I.~Y.~Aref'eva, A.~A.~Golubtsova and E.~Gourgoulhon,
``Thermalization of holographic Wilson loops in spacetimes with spatial anisotropy,''
Nucl. Phys. B \textbf{931}, 506-536 (2018)
[arXiv:1606.03995 [hep-th]].  
  
  
\bibitem{Lezgi:2021qog}
M.~Lezgi and M.~Ali-Akbari,
``Complexity and uncomplexity during energy injection,''
Phys. Rev. D \textbf{103}, no.12, 126024 (2021)
[arXiv:2103.05023 [hep-th]].

\bibitem{Ebrahim:2018uky}
H.~Ebrahim, M.~Asadi and M.~Ali-Akbari,
``Evolution of Holographic Complexity Near Critical Point,''
JHEP \textbf{09}, 023 (2019)
[arXiv:1811.12002 [hep-th]].

\bibitem{Amrahi:2020jqg}
B.~Amrahi, M.~Ali-Akbari and M.~Asadi,
``Holographic entanglement of purification near a critical point,''
Eur. Phys. J. C \textbf{80}, no.12, 1152 (2020)
[arXiv:2004.02856 [hep-th]].

\bibitem{Ebrahim:2020qif}
H.~Ebrahim and G.~M.~Nafisi,
``Holographic Mutual Information and Critical Exponents of the Strongly Coupled Plasma,''
Phys. Rev. D \textbf{102}, no.10, 106007 (2020)
[arXiv:2002.09993 [hep-th]].

\bibitem{Hashimoto:2018fkb}
K.~Hashimoto, K.~Murata and N.~Tanahashi,
``Chaos of Wilson Loop from String Motion near Black Hole Horizon,''
Phys. Rev. D \textbf{98}, no.8, 086007 (2018)
[arXiv:1803.06756 [hep-th]].
\bibitem{Eichten:1978tg}
E.~Eichten, K.~Gottfried, T.~Kinoshita, K.~D.~Lane and T.~M.~Yan,
``Charmonium: The Model,''
Phys. Rev. D \textbf{17}, 3090 (1978).


\bibitem{Bruni:2018dqm}
R.~C.~L.~Bruni, E.~Folco Capossoli and H.~Boschi-Filho,
``Quark-antiquark potential from a deformed AdS/QCD,''
Adv. High Energy Phys. \textbf{2019}, 1901659 (2019)
[arXiv:1806.05720 [hep-th]].

\bibitem{Andreev:2006ct}
O.~Andreev and V.~I.~Zakharov,
``Heavy-quark potentials and AdS/QCD,''
Phys. Rev. D \textbf{74}, 025023 (2006)
[arXiv:hep-ph/0604204 [hep-ph]].



\bibitem{Yang:2015aia}
Y.~Yang and P.~H.~Yuan,
``Confinement-deconfinement phase transition for heavy quarks in a soft wall holographic QCD model,''
JHEP \textbf{12}, 161 (2015)
[arXiv:1506.05930 [hep-th]].

\bibitem{Polchinski:2001tt}
J.~Polchinski and M.~J.~Strassler,
``Hard scattering and gauge / string duality,''
Phys. Rev. Lett. \textbf{88}, 031601 (2002)
[arXiv:hep-th/0109174 [hep-th]].

\bibitem{Witten:1998zw}
E.~Witten,
``Anti-de Sitter space, thermal phase transition, and confinement in gauge theories,''
Adv. Theor. Math. Phys. \textbf{2}, 505-532 (1998)
[arXiv:hep-th/9803131 [hep-th]].


  
\end{thebibliography}
\end{document}